\documentstyle[seceq, epsf]{ptptex}

\title{
${\rm CP}^{N-1}$ Models with a $\theta$ Term  and Fixed Point Action
}

\author{
Rudolf Burkhalter\footnote{E-mail: burkhalt@rccp.tsukuba.ac.jp}, 
 Masahiro Imachi$^{\ddagger}$\footnote{E-mail: imachi@sci.kj.yamagata-u.ac.jp},\\
  Yasuhiko Shinno$^{\diamond}$\footnote{E-mail: 
  shinno@dirac.phys.saga-u.ac.jp} and 
   Hiroshi Yoneyama$^{\diamond}$\footnote{E-mail: yoneyama@cc.saga-u.ac.jp}
}

\inst{
 Center for Computational Physics, University of Tsukuba, Tsukuba 
 305-8577, Japan\\
 Department of Physics, Yamagata University, Yamagata 990-8560, Japan$^{\ddagger}$\\
 Department of Physics, Saga University, Saga 840-8502, Japan$^{\diamond}$\\}

\recdate{
}

\abst{

The topological charge distribution $P(Q)$ is calculated for lattice ${\rm
CP}^{N-1}$ models. In order to suppress lattice cut-off effects,  we employ a
fixed point (FP) action. Through transformation of $P(Q)$,  we calculate the
free energy $F(\theta)$ as a function of the $\theta$ parameter. For N=4,
scaling behavior is observed for $P(Q)$ and  $F(\theta)$,  as well as the
correlation lengths $\xi(Q)$.  For N=2, however, scaling behavior is not
observed,  as expected. For comparison, we also make a calculation for 
the ${\rm CP}^{3}$ model with a standard action. We furthermore pay special
attention to the behavior of $P(Q)$ in order to investigate the dynamics of
instantons.  For this purpose, we carefully consider the  behavior of
$\gamma_{\it eff}$, which is an effective power of $P(Q)$ ($\sim
\exp(-CQ^{\gamma_{\it eff}})$), and reflects the local behavior of $P(Q)$
as a function of $Q$.  We study $\gamma_{\it eff}$ for two cases,
the dilute gas approximation based on the Poisson distribution of
instantons and the Debye-H\"uckel approximation of instanton 
quarks. In both cases,  we find
 behavior similar  to that  observed in numerical simulations.}

\begin{document}

\maketitle

\section{Introduction}
   It is  interesting  to 
  study the phase structure of  asymptotic free theories such as QCD 
  and the ${\rm CP}^{N-1}$ model.   
   Non-perturbative studies of the phase structure of such theories are 
necessary  in order to understand why effects of the topological term  
 ($\theta$ term) are suppressed  in Nature.  The $\theta$ term affects 
 the dynamics at low energy and is expected to lead to rich phase 
 structures.~\cite{rf:tH} 
  Actually, in the Z(N) gauge model, it has been shown by use of free energy 
 arguments that oblique confinement phases could emerge and that an 
 interesting phase structure may be realized.~\cite{rf:CR}
In this paper we are concerned with the dynamics of  the  $\theta$ 
  vacuum of ${\rm CP}^{N-1}$ models  with a topological  term,
   which have several  dynamical 
    properties in common   with QCD.   We believe that  study of the 
    two-dimensional 
    model will be useful in  acquiring information  about realistic 
    physics.  \par
     From the numerical point of view,  the topological term introduces
 a complex Boltzmann weight in the Euclidean 
lattice path integral formalism.  The complex nature of the weight   prevents
 one from straightforwardly
applying the standard algorithm used for Monte Carlo
 simulations.   This problem can be circumvented by Fourier-transforming 
 the topological charge distribution $P(Q)$.~\cite{rf:BDSL,rf:Wie}  
It is then necessary to 
 calculate $P(Q)$ as precisely as possible in order to reduce errors 
 in the expectation values of  physical operators as  functions of 
 $\theta$. The precise determination of  $P(Q) $ will allow us  to 
 obtain a  partition
  function and other quantities as  functions of $\theta$ 
  with high precision.  \par
   It is non-trivial  how lattice cut-off effects would emerge  through  the 
   introduction of a  topological term. 
    In the present paper, we  employ fixed point (FP) actions to study  
    this issue.~\cite{rf:HN}
    In the case of no topological term, the FP action is known to 
    significantly 
    suppress   lattice cut-off 
    effects for topological objects in ${\rm CP}^{N-1}$ 
    models.~\cite{rf:DFP,rf:BBHN,rf:Bu}
    In Ref.\cite{rf:BBHN}, a FP action for the ${\rm CP}^{1}$ model 
    (${\rm CP}^{1}$FP) was determined, and the 
    stability of instantons under minimization of the action was 
    investigated in detail. It was  also observed there that dislocations 
    can be  
    eliminated by adopting a FP charge as well as the FP action. 
    However,  the scaling behavior of the lattice topological susceptibility 
    $\chi_{t}$ was found to be strongly violated even after the
    dislocations are eliminated. 
    For the ${\rm CP}^{3}$ model with an FP action (${\rm CP}^{3}$FP),  
    contrastingly, impressive improvements have been  
    found;~\cite{rf:Bu} after the topological defects are removed,  clear 
    scaling behavior of $\chi_{t}$ is  observed. \par
      In the present paper, we study  the topological terms of ${\rm 
      CP}^{3}$ and ${\rm CP}^{1}$  models with a FP 
      action.  For comparison,  we also carry out  a calculation for 
      the ${\rm
    CP}^{3}$  model with the standard action (${\rm CP}^{3}$ST).
      The main purposes of the  paper are  the followings.
      \smallskip
      \begin{enumerate}
      \item To study the scaling  behavior of various quantities such as $P(Q)$,  the free energy, the 
      expectation value of the topological charge, the topological 
      susceptibility,  and correlation length as a  function of  $Q$.
      \item  To analyze  $P(Q)$ in detail   by considering  an effective 
      power  $\gamma_{\it eff}$  of $\ln P(Q)$.
      \end{enumerate} 
      \smallskip 
         The second of these purposes   is associated with the phase structure 
      of the model.
      In the very strong coupling region,  there exists a first-order transition 
      at $\theta = \pi$.~\cite{rf:Sei} In this 
      region $P(Q)$ is Gaussian, and its 
      volume dependence is given by  $P(Q)=C \exp (-\alpha/V Q^{2})$, where
      $C$ and $\alpha$ are $\beta$-dependent constants, $\beta$ being 
      the inverse coupling.  This 
      $1/V$-law of the exponent is associated with the existence of 
      the first-order phase transition at $\theta=\pi$.~\cite{rf:HITY,rf:IKY}
      When $\beta$ becomes larger for a fixed volume, $P(Q)$ has been found 
      to deviate from the Gaussian form.
      As a consequence, the singularity at  $\theta=\pi$ is no longer 
      visible. It is interesting to consider the  fate of the 
      first-order phase 
      transition.   In order to see how   $P(Q)$ varies, we define an effective 
      power $P(Q)\propto \exp (-C Q^{\gamma_{\it eff}})$,  which is defined 
      using  
      three adjacent values of $Q$ in  the whole range of $Q$.  We 
      investigate the  behavior of $\gamma_{\it eff}(Q)$ by systematically 
      varying $\beta$ and $V$. 
      For a fixed value of $\beta$, $\gamma_{\it eff}(Q)$ exhibits interesting 
      behavior as a function of the  topological charge density $Q/V$. 
      For a  small lattice size $L$,  $\gamma_{\it eff}(Q)$ approaches some 
      asymptotic value from below, while for large $L$ it  does so from 
      above.   For the  range investigated,   $\gamma_{\it eff}(Q)$ 
      is always  bounded  from below by 1.  As long  as the finite size 
      effects  are  not significant, $\gamma_{\it eff}(Q)$ is  between  
      1 and 2. Finite size effects are clearly seen in the behavior of 
      $\gamma_{\it eff}(Q)$ when it exceeds 2. \par          
          The second purpose stated  above is also associated with the dynamics of 
      instantons. We attempt  to extract some information about these 
      dynamics  from 
      $\gamma_{\it eff}$.  For this purpose, it is useful to  employ  
      two analytical  models. One is  that of a  dilute gas of 
      instantons  obeying the Poisson  distribution. For values of the 
      parameter corresponding to the very strong coupling region, the 
      Poisson distribution leads to a Gaussian form of $P(Q)$,
      but $P(Q)$ deviates from the Gaussian form as the coupling constant becomes 
      weaker. The other model  is  
      the Debye-H\"uckel (D-H) approximation of an instanton quark 
      gas.~\cite{rf:DM} This is based 
      upon an instanton quark picture,~\cite{rf:FFS} in which instanton quarks 
      interact weakly with each other.  
      For these two models,
      $P(Q)$ is calculated from the partition function $Z(N_{+}, N_{-})$,
      which is the  probability  to generate $N_{+}$ instantons and $N_{-}$
      anti-instantons. The quantities $\gamma_{\it eff}(Q)$  for the two 
      models exhibit behavior similar to 
      that found in  Monte Carlo simulations. Our conclusion is that the 
      distribution $P(Q)$ generated by 
      Monte Carlo simulations are not inconsistent with the 
      dilute gas approximation.  \par
    In the following section, we briefly summarize the notation and 
 the algorithm of the complex action calculation.
In section~\ref{sec:CP3results} we present the results for ${\rm
CP}^{3}$FP,  and in section~\ref{sec:CP1results} we compare them with
those of ${\rm CP}^{1}$FP and ${\rm CP}^{3}$ST. In 
section~\ref{sec:DH}, 
we summarize and discuss analytical results for the Debye-H\"uckel 
approximation of an instanton quark gas.
We also compare values of $\gamma_{\it eff}$ from numerical simulations 
with those  
obtained from the Debye-H\"uckel model and from the Poisson
distribution. A summary is given in section~\ref{sec:Remarks}. \par

\section{Formulation} 
\label{sec:Form}
\subsection{Definition and algorithm}
 The action with the $\theta$ term is defined by
  \begin{equation}
 S_\theta= S -  i \theta \hat{Q},
 \end{equation} 
 where $S$ is a lattice action of the  ${\rm CP}^{N-1}$ model. 
 Among various definitions of the topological charge, we here choose 
 the geometrical definition.~\cite{rf:BeLu}
 The topological charge $Q$ is counted by $A_{\Box}$ as
 \begin{equation}
 Q=\frac{1}{2\pi}\sum_{\Box}A_{\Box}
	\label{eqn:gq}
 \end{equation}
 in the updating process, where the plaquette contribution $A_{\Box}$ is given
 by
 \begin{equation}
 A_{\Box}=\frac{1}{2}\sum_{\mu,\nu}\{ A_{\mu}(n)+A_{_\nu}(n+\mu)-A_{\mu}(n+\nu)
  -A_{\nu}(n+\mu)\} \epsilon_{\mu\nu}  \;\; {\rm mod}  \; 2\pi.
 \end{equation}
 Here  $\exp(iA_{\mu}(n))\equiv z^{\dag}(n)z(n+\mu)/|z^{\dag}(n)z(n+\mu)|$, 
 i.e., $A_{\mu}(n)\equiv\arg(z^{\dag}(n)z(n+\mu))$. The link variables 
 $A_{\mu}(n)\equiv\arg(z^{\dag}(n)z(n+\mu))$ satisfy 
 \begin{equation}
 A_{\mu}(n)\in[-\pi,\pi]. \nonumber
 \end{equation} 
  In order to avoid the complex Boltzmann weight, we adopt an 
  algorithm by which the partition function is given by the Fourier 
  transform of the topological  charge distribution  $P(Q)$
  \begin{equation}
      	Z(\theta)=\sum_{Q}e^{i \theta Q} P(Q).
      	\label{pa}
      \end{equation}   
 The distribution  $P(Q)$ is given in terms of  the real Boltzmann weight 
 as 
 \begin{equation}
    	P(Q)=\int [dz d{\overline z}]^{Q}e^{-S}/\int [dz d{\overline z}]e^{-S},
    	\label{pq}
    \end{equation}
    where $[dz d\overline{z}]^{Q}$ is the constrained   measure on   which the 
   value of the  topological charge given in Eq. (\ref{eqn:gq})  is restricted to  $Q$. $P(Q)$ 
   is normalized as  $\sum_{Q} P(Q)=1$.\par
 The expectation value of an  operator $O$  as a function of $\theta$  is given by 
 \begin{equation}
 	\langle O (\theta) \rangle = \sum_{Q}\exp[i \theta Q] P(Q) \langle 
 	O \rangle_{Q} / Z(\theta),	
 	\label{op}
 \end{equation}
 where 
 \begin{equation}
 	\langle  O \rangle_{Q}= \int [dz d{\overline z}]^{Q} O e^{-S}/ \int [dz d{\overline z}]e^{-S}.
 	\label{oq}
 \end{equation}
    We calculate $P(Q)$ by updating   configurations through the  combined   use  of 
 the overrelaxation algorithm and the Metropolis algorithm for ${\rm 
 CP}^{3}$, 
 while only the Metropolis algorithm  is applied for   ${\rm CP}^{1}$.  From the generated   
 configurations, the topological charge is calculated according  to 
 Eq. (\ref{eqn:gq}).
 Since each  function $P(Q)$  under consideration   rapidly falls off, 
 it is convenient to restrict the range of $Q$ for  a single Markoff 
 chain. We use the set method~\cite{rf:BBCS,rf:KSC} by which an entire range of $Q$ is 
 divided into sets $S_{i}, i=1, 2, 3,\cdots$. 
 Typically,  each of the sets $S_{i}$
      consists of 4 bins,  $Q=3i-3, 3i-2, 3i-1$ and  $3i$,  so that the adjacent set 
      overlaps at the edge bin of the set,  $Q=3k\, (k=1,2,3,\cdots)$.  
 Depending on $\beta$, the volume and also  $N$, it is sometimes more
 convenient to use  a wider range of bins for a  set to   save 
 computation  time,  
   as well as  to allow for  better use of the    trial function method 
   described below. \par
 In order to generate  configurations more efficiently,  
 an effective action is used.~\cite{rf:KSC} This action is modified by 
 adding a trial function $P_{t}(Q)$ according to 
  \begin{equation}
   S_{\it eff}=S-\ln P_{t}(Q).
 	\label{efa}
 \end{equation}
 The form of $P_{t}(Q)$ is  chosen   to be 
 \begin{equation}
 	P_{t}(Q)\propto \exp (-\alpha Q^{\gamma}),
 	\label{tf}
 \end{equation}
 where $\alpha$ and  $\gamma$ are adjusted  so that $P(Q)$ becomes 
 almost flat,  in order to reduce errors. The power $\gamma$ is often chosen to 
 be 2.0 (the Gaussian case).  
\bigskip
\subsection{Fixed point action}
   
The idea of using a FP action in asymptotically free theories in order to
remove lattice artifacts was introduced by Hasenfratz and
Niedermayer.~\cite{rf:HN} 
The action is defined at the fixed point of a renormalization
group transformation at $\beta=\infty$ and is perfect, i.e. without
any discretization errors, on the classical level.  Although on the quantum
level the FP action is not perfect, it is in practice powerful with respect
to removing lattice defects. 
For a finite coupling constant $\beta$, the action $S$  is given 
by
  \begin{equation}
   S=N\beta  A_{{\rm FP}},
 \end{equation}
where $A_{{\rm FP}}$ is the FP action.  $A_{{\rm FP}}$ is determined by 
a block transformation 
at $\beta=\infty$  to be 
  \begin{equation}
   A_{{\rm FP}}(\zeta)={\rm min}_{z}[A_{{\rm FP}}(z)+T(\zeta, 
   z)], 
 	\label{fps}
 \end{equation}
 with the transformation kernel $T$ for  block transformation from ${\rm 
 CP}^{N-1}$ spins $z$ 
 to block spins $\zeta$ (see Ref.~\cite{rf:HN} for details).
It has been shown that lattice defects are
invisible up to fairly small coupling constants that  correspond to
correlation lengths of only a few units of the lattice spacing.  Because of
these benefits we use the FP action to study the ${\rm CP}^{3}$ model with
a topological term.
\par
  
For numerical simulations,  a parameterized form of the FP action is required. 
In previous studies, a large number of coupling constants have been used
for parametrization.  In Ref.~\cite{rf:Bu}, for example, ${\rm
CP}^{3}$FP has been parameterized using 32 coupling constants in order to
realize high precision. In order to reduce the computational effort 
required to carry out simulations with a FP action, we constructed for the 
present work a simpler
parametrization using the same method and the same set of configurations as
in Ref.~\cite{rf:Bu}. We were able to reduce the number of coupling
constants from 32  to 9, while only increasing the average relative
deviation between the minimized and the parameterized FP action from 0.4\%
to 0.6\%.  All the couplings of the parametrization are limited to a short
range and lie within one plaquette. We list the coupling constants of the 
simpler
parametrization in Table~\ref{table:couplings},  where we use the same
numbering scheme as in Ref.~\cite{rf:Bu}. \par

For ${\rm CP}^{1}$FP, we employ the same set of coupling constants as  in
Ref.~\cite{rf:HN} (24 types of coupling constants). \par

\begin{table}
\caption{Coupling constants of the ${\rm CP}^{3}$FP action used in this work.}
\label{table:couplings}
\begin{center}
\begin{tabular}{cc|cc|cc} 
\hline 
\hline
No. in Ref.~\cite{rf:Bu} & Coupling  & 
No. in Ref.~\cite{rf:Bu} & Coupling  & 
No. in Ref.~\cite{rf:Bu} & Coupling  \\
\hline
1  &  0.61884 &  2  & $-0.04879$ &  4  &  0.19058 \\  
7  &  0.00645 &  11 &   0.02984  &  17 & $-0.12433$ \\  
19 &  0.04034 &  20 &   0.14477  &  23 &  0.01852 \\  
\hline
\end{tabular}
\end{center}
\end{table}

\subsection{Measurements}

The free energy density $F(\theta)$ is obtained from the partition 
function of Eq.~(\ref{pa}) through the relation
  \begin{equation}
 	F(\theta)=-\frac 1 V \ln Z(\theta),
 	\label{fn}
 \end{equation}   
where $V=L^{2}$,  and $L$ is a dimensionless lattice extension. The
expectation value of the topological charge is defined as
  \begin{equation}
 	\langle Q \rangle_{\theta} = - (-i) \frac{dF(\theta)}{d\theta}.
 	\label{qn}
 \end{equation}   

The correlation length $\xi(Q)=1/m(Q)$ for a fixed topological charge sector
is obtained from two point functions of $P={\overline z}\otimes z$
projected to zero momentum. It is extracted by analyzing their long 
distance fall-off of the form
\begin{equation}
	\langle P(0)P(\tau)\rangle_Q =A\left( \exp\left(-m(Q) \tau\right) 
	+\exp\left(-m(Q)\left( T-\tau\right)\right)\right),
	\label{tw}
\end{equation}
where $T$ is the  extent of the lattice in the time direction. \par %

\section{Numerical results for ${\rm CP}^{3}$FP}
\label{sec:CP3results}

\subsection{${\rm CP}^{3}$FP without a topological term}
\label{sec:CP3theta0}

Before presenting the results of our simulations using the set method,  we briefly
discuss numerical results of a series of standard simulations using the FP
action for the ${\rm CP}^{3}$ model. There are two reasons why it is useful
to perform these additional simulations. First, it has been observed~\cite{rf:Bu}
 that use  of a FP action leads to good scaling
behavior of the topological charge  even with  a coarse
lattice. Since in the present work we use a simpler parametrization of the
FP action, we have to check that these  good scaling properties are kept for
this new action. The second reason is that from simulations without a
topological term basic information about correlation lengths is
obtained. This is needed to answer fundamental questions such as which
couplings should be considered as strong and which as weak and  which values
of the lattice size $L$ correspond to a large lattice in physical units and
which to a small one. Values of the correlation length are also required in
order to allow for investigation  of scaling behavior. In the following sections
it is understood, if not otherwise stated, that the correlation length
is that  determined from simulations without a topological term.

\begin{table}[htb]
\caption{Results of simulations with ${\rm CP}^{3}$FP without a topological
term.}
\label{table:resCP3FP}
\begin{center}
\begin{tabular}{ccr@{.}lr@{.}lr@{.}lr@{.}lr@{.}lr@{.}l}
\hline
\hline
$\beta$ & L & \multicolumn{2}{c}{$\xi$}  &
\multicolumn{2}{c}{$L/\xi$} &
\multicolumn{2}{c}{$\chi^t_{\rm ST}$} &
\multicolumn{2}{c}{$\chi^t_{\rm FP}$} &
\multicolumn{2}{c}{$\chi^t_{\rm ST}\xi^2$} &
\multicolumn{2}{c}{$\chi^t_{\rm FP}\xi^2$} \\
\hline
0.5 & 10 &  0&4515(61) & 22&15(30)  & 0&05654(11)   & 0&05394(11)   &
   0&0115(3)    & 0&0110(3)  \\
1.0 & 10 &  0&7019(18) & 14&247(36) & 0&044047(89)  & 0&041471(84)  &
   0&0217(1)    & 0&0204(1)  \\
1.5 & 10 &  1&0790(12) &  9&268(10) & 0&029886(60)  & 0&027989(57)  &
   0&0348(1)    & 0&0326(1)  \\
3.0 & 32 &  7&657(35)  &  4&179(19) & 0&0014829(25) & 0&0013928(23) &
   0&0869(8)    & 0&0817(8)  \\
3.0 & 46 &  7&114(27)  &  6&466(24) & 0&0016151(30) & 0&0015262(28) &
   0&0817(6)    & 0&0772(6)  \\
3.2 & 60 &  9&656(84)  &  6&214(54) & 0&0008848(18) & 0&0008400(17) &
   0&0825(14)   & 0&0783(14) \\
3.4 & 82 & 13&08(17)   &  6&267(81) & 0&0004786(16) & 0&0004549(16) &
   0&0819(25)   & 0&0778(20) \\
\hline
\end{tabular}
\end{center}
\end{table}

In Table~\ref{table:resCP3FP} we list the  run-parameter values used and 
the main results of
simulations without a topological term. We chose three values of $\beta$
(0.5, 1.0 and 1.5) in the strong coupling region,  which correspond to
correlation lengths of one lattice spacing or less. In the weak coupling
region,  we also chose three $\beta$ values,  corresponding to correlation
lengths of more than seven lattice spacings. Calculations were  performed on
square lattices of size $L$, chosen such that the condition $L/\xi\geq 6$ is
fulfilled in order to avoid finite size effects.
The condition $L/\xi\geq 6$ is chosen to be approximately the same as that in
Ref.~\cite{rf:Bu}. With  this value, a clear scaling plateau of the
topological susceptibility (as  discussed below) has
been found.  There are finite size effects even for this
physical size, but the effects are sufficiently small  to see the scaling
behavior for different values of $\beta$ ( $\beta= 3.0, 3.2$ and 3.4 ), where
the effects are  expected to be similar.
The only exception to this
is the run performed with  $\beta=3.0$ and $L=32$. The difference 
between the  results of this
particular run and  that  on a larger lattice clearly demonstrates the existence
of finite size effects on smaller lattices, which is in accordance 
with 
similar observations in Ref.~\cite{rf:Bu}. For each run, 
several million sweeps were carried out.\par
During the simulations,  we measured the topological charge using two
definitions. One is the standard geometrical charge $Q_{\rm ST}$ defined in
Eq.~(\ref{eqn:gq}). In addition,  we  measured the  FP topological charge
$Q_{\rm FP}$, which is  defined as in Ref.~\cite{rf:Bu} on a first finer
level of a
multigrid,  calculated using a parametrization of the FP field. The
topological susceptibility is obtained through the relation
\begin{equation}
\chi^t = \frac{\langle Q^2 \rangle}{V},
\end{equation}
for which numerical values are given in Table~\ref{table:resCP3FP}. As in
Ref.~\cite{rf:Bu}, we find that $\chi^t_{\rm FP}$ is somewhat smaller
than $\chi^t_{\rm ST}$.  In order to study scaling,  we consider the
dimensionless combination
$\chi^t \xi^2$, whose values  are  also listed in Table~\ref{table:resCP3FP}.
We find this   quantity to be approximately constant for both 
cases:
  $\chi^t_{\rm ST}\xi^2 \approx 0.082$  and $\chi^t_{\rm FP}\xi^2 \approx 0.077$
  for the three weak couplings.  These values  are 6 to
  12 $\%$ larger  than  the results in  Ref.~\cite{rf:Bu} but
  display a  clear scaling plateau  in the   same region of $\xi$.
This confirms that the new simpler parametrization of the FP action, even 
without use of the
  FP topological charge,  has the same good scaling properties.
  This becomes even clearer when one compares the behavior with the
  results   for  the standard action case,~\cite{rf:HM} where this  plateau has 
never
  been found. Since a combined  use of $Q_{\rm FP}$ and the set method
  requires   too much  computation time, in the simulations discussed 
  below, we   used  $Q_{\rm ST}$ for 
measurements of
  the   topological charge.
  %
\subsection{Scaling behavior of $P(Q)$ and $\xi(Q)$}
  In this subsection we present the results for  $P(Q)$ and $\xi(Q)$
   and discuss  the scaling behavior of these quantities.

\subsubsection{$P(Q)$}
\label{sec:CP3PQ}
Calculations of the topological charge distribution $P(Q)$ using the set
method were  performed for various values of the coupling constant $\beta$
and for various lattice volumes $V=L^2$. An overview is given in
Table~\ref{table:parCP3FP},  where we also indicate the range $Q_{\rm
min}$--$Q_{\rm max}$ for which $P(Q)$ is calculated in each case.

\begin{table}
\caption{Parameter values used in simulations of ${\rm CP}^{3}$FP with 
the set method.}
\label{table:parCP3FP}
\begin{center}
\begin{tabular}{cr@{:}lr@{:}lr@{:}lr@{:}lr@{:}l}
\hline
\hline
$\beta$ & \multicolumn{10}{c}{$L$ : $Q_{\rm min}$--$Q_{\rm max}$} \\
\hline
0.5 &  6 \,&\, 0--12  & 10 \,&\, 0--18   \\
1.5 &  8 \,&\, 0--18  & 12 \,&\, 0--24 & 24 \,&\, 0--39  \\
3.0 &  4 \,&\, 0--4   &  6 \,&\, 0--12 &  8 \,&\, 0--18  & 12 \,&\, 0--30 &
       18 \,&\, 0--54  \\
     & 24 \,&\, 0--58  & 28 \,&\, 0--3  & 32 \,&\, 0--94  & 34 \,&\, 0--3 &
       38 \,&\, 0--27  \\
     & 42 \,&\, 0--3   & 46 \,&\, 0--33 & 50 \,&\, 0--3   & 56 \,&\, 0--3 &
       96 \,&\, 0--60 \\
3.4 & 22 \,&\, 0--18  & 44 \,&\, 0--27 & 58 \,&\, 0--36  \\
\hline
\end{tabular}
\end{center}
\end{table}

We carried out  an extensive study of the lattice size dependence 
with  $\beta=3.0$, 
which we have seen in the previous subsection  already lies in the scaling
region. Lattice sizes were  chosen in a wide range,  beginning from very
small ones with $L/\xi < 1$ up to large ones with $L/\xi \approx 13$.  This
study was  supplemented by additional runs using the  larger value of 
$\beta=3.4$. This was done in order to make more detailed checks of scaling.
 Lattice sizes
here were  chosen so that the values of $L/\xi$ are approximately the 
same as those used in the $\beta=3.0$ case, resulting in
the identifications $22
\Leftrightarrow 12$, $44 \Leftrightarrow 24$ and $58 \Leftrightarrow 32$.
In order to study the differences between the weak coupling region and the
strong coupling region, some additional simulations were also performed 
with 
the strong couplings $\beta=0.5$ and 1.5. Since for  strong coupling the
correlation length is too short, the physical lattice size $L/\xi$
cannot be taken small. For these studies we have therefore been
restricted to large physical lattice sizes in the range
$13 \leq L/\xi \leq 22$ for
$\beta=0.5$ and $7 \leq L/\xi \leq 22$ for $\beta=1.5$.

The statistics of simulations with  $\beta=3.0$ consisted  of the order of
10--20 million sweeps per set for the cases in which  several sets 
were 
employed. Higher statistics were obtained, with approximately  50 million 
sweeps,   for
simulations with $L=34$, 42, 50 and 56,  for which only one set was  employed
in order to study the behavior around $Q=0$. The highest statistics,  with
150 million sweeps,  were obtained  for $L=46$ in the first set,  $Q=0$ --
3. Increasing the statistics for this case, particularly in the first 
set,  was
crucial to reduce  errors in  the free energy density $F(\theta)$ for 
$\theta$ close to $\pi$. Statistics for  $\beta=0.5$, 1.5 and 3.4 
consisted  of
about 2 million sweeps per set.
  For error analysis,  we employ
the jack-knife method by dividing the runs into approximately 100 subsets.
Since autocorrelation times  for $Q$ are at most a few of tens of time 
units,  
 2 million sweeps were sufficient to  make the
measurements  independent and the errors  reliably estimated.
\begin{figure}
\vspace{-5mm}
     \epsfxsize=12cm
     \centerline{\epsfbox{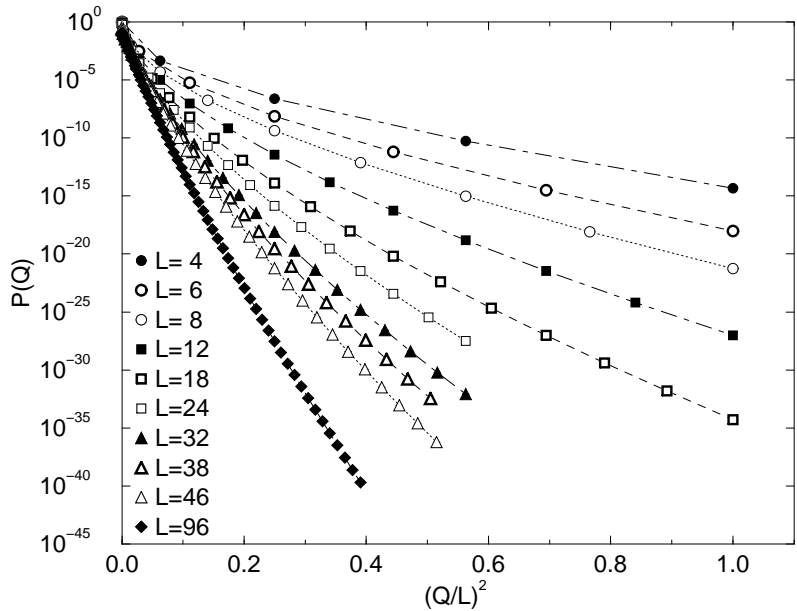}}
\vspace{-5mm}
\caption{Probability distribution of the topological charge for ${\rm
CP}^{3}$FP with  $\beta=3.0$ and various values of $L$. Only the results with $Q/L
\leq 1$ are plotted. The size  of error bars is such that they cannot 
be seen  at this scale of plotting. The curves  connecting the data points are
included as a guide to the eye.}
\label{fig:CP3PQ}
\end{figure}

In Fig.~\ref{fig:CP3PQ} we plot the measured topological charge
distribution $P(Q)$ with  $\beta=3.0$ for various lattice sizes $L$. For
convenience and to present data in a compact manner,  we normalize the
topological charge with the lattice size $L$. Even a rough inspection of
the figure  reveals that the data do not exhibit Gaussian behavior,
which would be represented by straight lines with this method of plotting. In
particular,  the data for small lattice sizes exhibit a clear 
curvature,  while they
tend to straighten out to some extent as the lattice size increases. However,
fits with a Gaussian form turn out to be extremely poor for all the cases.

\begin{figure}
\vspace{-5mm}
     \epsfxsize=12cm
     \centerline{\epsfbox{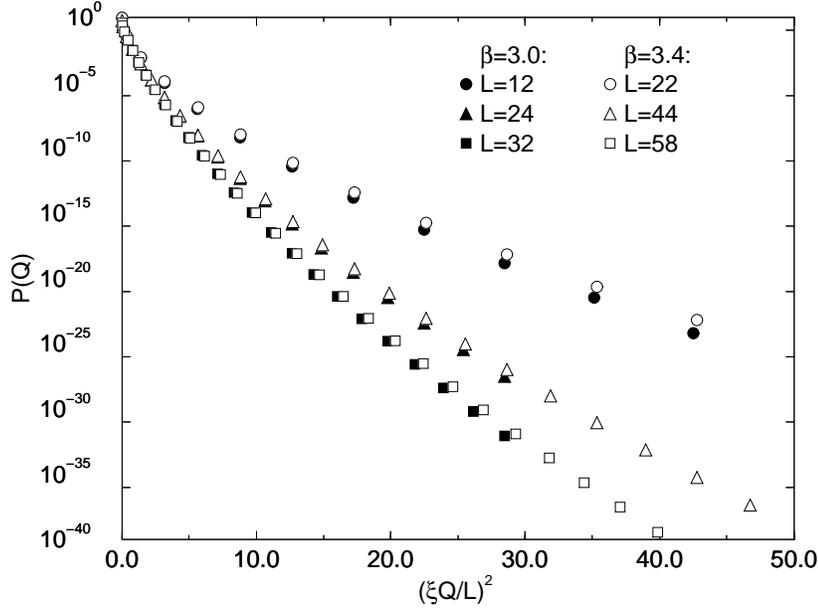}}
\vspace{-5mm}
\caption{Scaling test of the charge probability distribution for ${\rm
CP}^{3}$FP at two values of the coupling and several volumes.}
\label{fig:CP3PQScal}
\end{figure}

  In Fig.~\ref{fig:CP3PQScal} we
compare the charge distribution obtained with  the two weak couplings
$\beta=3.0$ and 3.4 for a scaling check of $P(Q)$. For such a comparison to 
be possible,  we have now to
normalize the charge with the physical lattice size $L/\xi$,  for which we use
measured values of the correlation length $\xi$. We find that the  data at the
two couplings exhibit the same behavior and lie roughly on the same 
line.  The  remaining differences are  most visible towards larger values of the
topological charge. 
This may be explained by  the following two reasons. One is that the 
 physical lattice sizes can differ slightly.    Values of  $L/\xi$ are 
  1.687(6) (L=12), 3.37(1) (L=24), 4.50(2) (L=32) for $\beta=3.0$ and 
  1.68(2) (L=22), 
  3.36(4) (L=44),  4.43(6) (L=58) for $\beta=3.4$.  
   The other is that  $\xi$
is chosen  from Table~\ref{table:resCP3FP} for 
  each $\beta$ .  We take $\xi$ to be 7.114(27) for $\beta=3.0$ and 
  13.08(17) for $\beta=3.4$. If  the measured  values of  $\xi$  were  used  for each  
  $L$,   one would expect at most a 10 $\%$ 
  difference among  the values of $\xi$  in this range of $L$, but the quality of 
  the  agreement of the values of $P(Q)$ in the  figure does not change.
  In a strict sense, we are not able to
obtain rigorous scaling behavior, but the quality of the agreement of
the two different values of $\beta$ is still remarkably good compared to 
the other two models, 
${\rm CP}^{1}$FP  and ${\rm CP}^{3}$ST. This point  is  discussed in
detail in the following section (see  Fig.~\ref{fig:CP1PQ}).
For an argument of more rigorous  scaling,  we  need to use measured 
$\xi$ for each $L$ and  use $Q_{{\rm
FP}}$ instead of $Q_{{\rm ST}}$, but  here we think the results in
  Fig.~\ref{fig:CP3PQScal} are sufficient for the present  discussion.
\subsubsection{Correlation length}
\label{sec:CP3XI}
\begin{figure}
\vspace{-2cm}
     \epsfxsize=14cm
     \centerline{\epsfbox{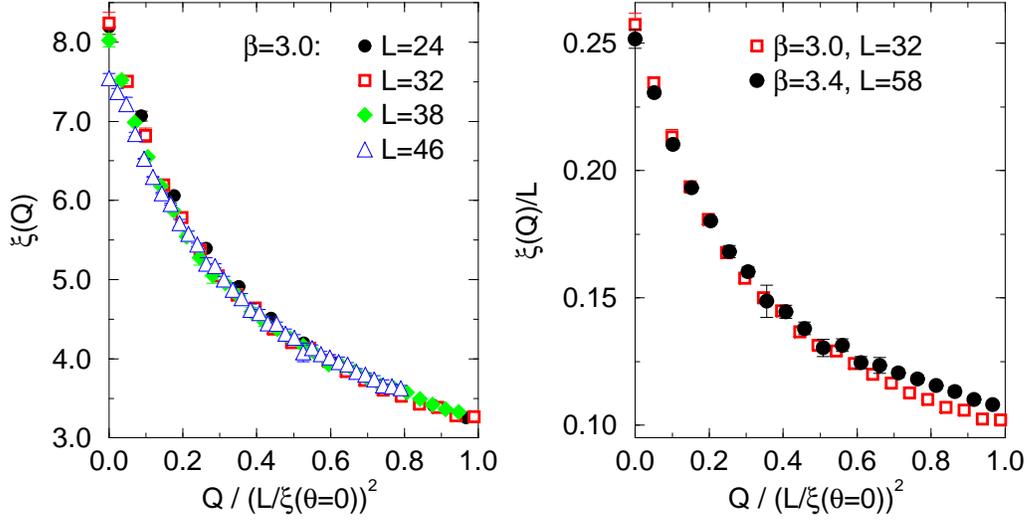}}
\vspace{-5mm}
\caption{Fixed charge sector correlation length as a function of the
physical charge density for ${\rm CP}^{3}$FP. In the left panel we compare
various lattice sizes at $\beta=3.0$ and in the right panel we show scaling
between $\beta=3.0$ and 3.4.}
\label{fig:CP3Qxi}
\end{figure}

Simulations with the set method allowed us  to determine the dependence of the
correlation length on the topological charge. Correlation functions can be
measured separately for different topological sectors,  and correlation
lengths $\xi(Q)$ can be extracted from their fall-off at large time
separations.
The results are shown in Fig.~\ref{fig:CP3Qxi},  where we plot correlation
lengths as functions  of the charge density.  We normalize the charge
density by  the
correlation length $\xi(\theta=0)$ obtained from simulations without
a topological term discussed in Sec.~\ref{sec:CP3theta0}.\par
In the left panel of Fig.~\ref{fig:CP3Qxi},  we plot  the data obtained 
with 
$\beta=3.0$ and different lattice sizes. As  expected, the data fall
on a universal curve  if plotted as  functions of the topological charge
density.  We find a decrease of the correlation length with increasing
topological charge density. The reason for this is that with an increasing
number of topological objects within the same volume,  the configuration has
to become less ordered.  At the point where the topological charge density
$Q/(L/\xi(\theta=0))^{2}$ is equal to 1, the correlation length $\xi(Q)$ has
dropped to less than half of its value at zero topological charge.
In the right panel of Fig.~\ref{fig:CP3Qxi} we compare representative
data at two different values of the coupling constant.  We see clear
scaling behavior of $\xi(Q)/L$. For smaller values of $\xi(Q)$, on the
other hand, we have observed violations of the scaling property. The curve
for $\beta=3.0$ starts to deviate at $Q/(L/\xi(\theta=0))^{2} \approx 0.7$
where
$\xi\approx 4$. This,  in turn,  indicates that even a use of the FP action is
unable to suppress the lattice defects for  $\xi \leq 4$, which is in
agreement with similar previous  observations.~\cite{rf:Bu,rf:BK} \par
 From these scaling checks,  we conclude that
the observed behaviors of $P(Q)$  and $\xi(Q)$ for  $\beta=3.0$ 
represent   continuum
properties and are not caused by lattice artifacts. These are 
compared with the results for ${\rm CP}^{1}$FP  and ${\rm CP}^{3}$ST actions
  in the following section.

\subsection{Free energy density and expectation value of the topological charge}
\label{sec:CP3FTH}
\begin{figure}
\vspace{-5mm}
     \epsfxsize=12cm
     \centerline{\epsfbox{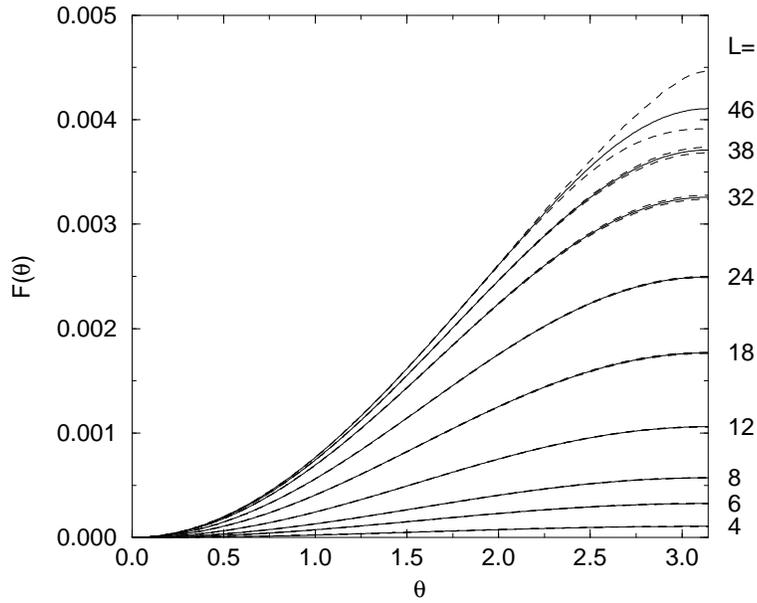}}
\vspace{-5mm}
\caption{Free energy density $F(\theta)$ for ${\rm CP}^{3}$FP for
$\beta=3.0$ and several lattice sizes $L$.}
\label{fig:CP3freeen}
\end{figure}

In Fig.~\ref{fig:CP3freeen} we show results for the free energy density
obtained with Eq.~(\ref{fn}) for $\beta=3.0$ and several lattice sizes
$L$.  Central values are indicated by solid lines, while the one-sigma
error bands, determined with a jack-knife analysis, are represented by
dashed lines. We obtain smooth curves, and no ``flattening'' is observed, as
in some other works.~\cite{rf:IKY,rf:Sch,rf:PS} The only
exception is the simulation with $L=96$ (not
shown in the figure), where the calculation of $F(\theta)$ breaks down at
$\theta \approx 1$, because the partition function turns out to be negative
(while still consistent with zero within the error bars). 
Before this breakdown, the error bars become very large.

Note that the physical free energy densities for the corresponding lattice
sizes of $\beta=3.0$ and $\beta=3.4$ fall on the same curve, which is a
consequence of the scaling of $P(Q)$ shown in Fig.~\ref{fig:CP3PQScal}.
We observe a strong dependence of the free energy density on the lattice
size. $F(\theta)$ increases with increasing $L$ and for $\theta < 1$,
already, nearly asymptotic values are obtained for our largest lattice sizes.
This is, however, not yet the case for $\theta$ close to $\pi$. This has
the consequence that curves become steeper as the volume increases. Does this
mean that $F(\theta)$ develops a peak at $\theta=\pi$, which would be an
indication of a phase transition?

\begin{figure}
\vspace{-5mm}
     \epsfxsize=12cm
     \centerline{\epsfbox{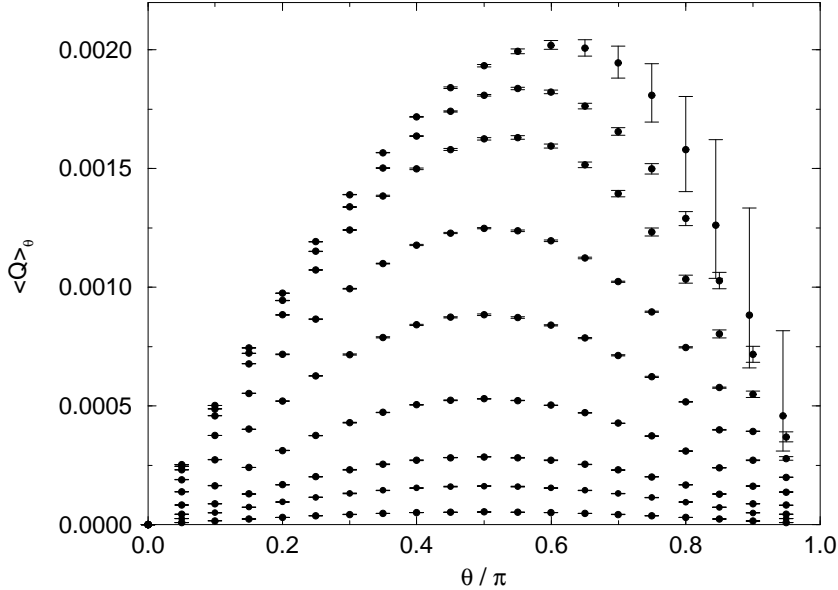}}
\vspace{-5mm}
\caption{Expectation value of the topological charge $\langle Q
\rangle_{\theta}$ for ${\rm CP}^{3}$FP with $\beta=3.0$ and
several lattice sizes $L$. Results increase from $L=4$ (data points at the
bottom) to $L=46$ (data points at the top).}
\label{fig:CP3qexp}
\end{figure}

\begin{figure}
\vspace{-5mm}
     \epsfxsize=12cm
     \centerline{\epsfbox{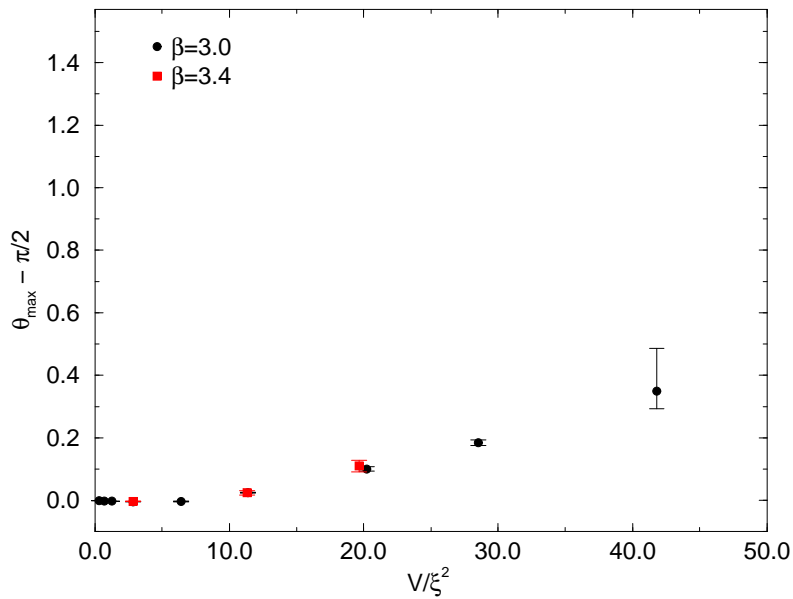}}
\vspace{-5mm}
\caption{Angle $\theta_{\rm max}$ of the maximal topological charge for ${\rm
CP}^{3}$FP as a function of the physical lattice volume.}
\label{fig:CP3thmax}
\end{figure}

To answer this question, we consider the expectation value of the
topological charge obtained with Eq.~(\ref{qn}). We show results obtained
for $\beta=3.0$ in Fig.~\ref{fig:CP3qexp}. As seen there, 
$\langle Q \rangle_{\theta}$
vanishes both for $\theta=0$ and $\pi$. In between these values, it has 
a peak, which we
observe to move slowly away from $\theta=\pi/2$, obtained at small lattices,
towards $\pi$. Figure~\ref{fig:CP3thmax} displays the volume dependence of
$\theta_{\rm max}$, defined to be the position of the peak of $\langle Q
\rangle_{\theta}$. This clearly shows how $\theta_{\rm max}$ moves
away from $\pi/2$. It is, however, still far from $\pi$ even with our
largest lattice, and it cannot be conclusively determined from our data 
where the infinite volume limit of
$\theta_{\rm max}$ is.

\subsection{ Effective power $\gamma_{\it eff}$ of $P(Q)$}
\label{sec:CP3GEFF}
We are now in a position to investigate the behavior of the charge
distribution in more detail without having to be concerned about discretization
effects. This is important for obtaining information about the 
underlying  physics.
  Having observed that $P(Q)$ cannot be fit with a simple Gaussian,
we turn to a closer examination of its local properties. To this end, we
calculate the effective power $\gamma_{\it eff} =
\gamma_{\it eff}(Q)$ defined by assuming that $P(Q)$ behaves at the three
adjacent charges $Q$, $Q+1$ and $Q+2$ as the function
\begin{equation}
P_{\rm local}(Q) = A \exp(-BQ^{\gamma_{\it eff}}).
\label{eqn:gammaeff}
\end{equation}
Since three values of the charge distribution are used as inputs and the
function in Eq.~(\ref{eqn:gammaeff}) has three free parameters, the latter
can be calculated algebraically without involving a fit. Trivially,
the Gaussian distribution takes the value $\gamma_{\it eff}=2$, independent of
$Q$.\par
\begin{figure}
\vspace{-15mm}
     \epsfxsize=12cm
     \centerline{\epsfbox{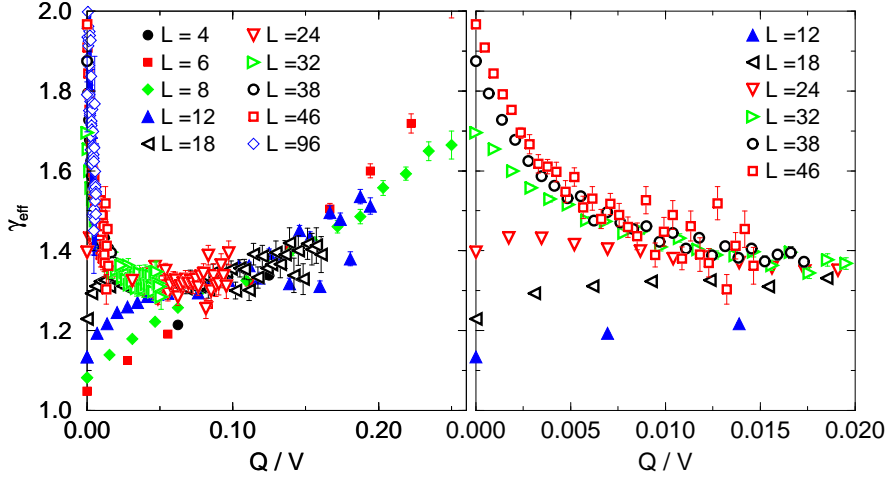}}
\vspace{-15mm}
\caption{Effective power of the probability distribution for ${\rm
CP}^{3}$FP with $\beta=3.0$ and various lattice sizes. The left panel shows
all results up to a charge filling fraction of $1/4$, while the right panel
is an enlargement of the region with small filling fraction.}
\label{fig:CP3geff}
\end{figure}

Results for $\gamma_{\it eff}(Q)$ with $\beta=3.0$ are displayed in
Fig.~\ref{fig:CP3geff} as a function of the charge density or charge
filling fraction $Q/V$ of the lattice. We plot the same data twice, first
showing all results up to large filling fractions in the left panel, and
then zooming in to the region of small filling fractions in the right
panel,
  where data for small (4, 6 and 8) and large (96) lattice sizes
are omitted for  clearer visibility.
We observe that for almost all the combinations of lattice size $L$ and
topological charge $Q$, the values of $\gamma_{\it eff}$ differ considerably
from 2. This explains why we could not fit $P(Q)$ with a
Gaussian. In fact,  $P(Q)$ is not even well described locally by a Gaussian
for most of the cases. We find, however, that the values of 
$\gamma_{\it eff}(Q)$
are bounded between the two limits 1 and 2. The values found at $Q=0$ 
depend on the lattice size. When the topological charge is
increased, we find that results from different lattice sizes seem to
approach a common value of $\gamma_{\it eff} \approx 1.3$. This common
value is reached, at the latest, at a filling fraction of around $1/10$.  An
interesting point to observe is that the approach to this common value is
from two different sides, depending on the lattice size. Data for $L=12$, for
example, start with $\gamma_{\it eff} = 1.13$ at $Q=0$ and increase with
increasing charge. As an example of different behavior, data for $L=38$
start with $\gamma_{\it eff} = 1.88$ at $Q=0$ and decrease with increasing
charge. The lattice size at which increasing behavior changes to 
decreasing behavior is approximately $L = 24$ (which corresponds to $L/\xi
\approx 3.5$).\par
If we increase the topological charge to filling fractions larger than
about $1/10$, the values of $\gamma_{\it eff}$ move away from 1.3 and increase
in a universal manner. This can be easily understood to be an effect of 
approaching the limit of the maximal number of topological objects
(instantons) that can be placed on a lattice. The probability of such
densely packed lattice configurations is decreased, which, in turn, leads to
an enhancement of $\gamma_{\it eff}$. This point will be investigated 
in section~\ref{sec:DH} by studying analytical models.\par
\begin{figure}
\vspace{-5mm}
     \epsfxsize=11cm
     \centerline{\epsfbox{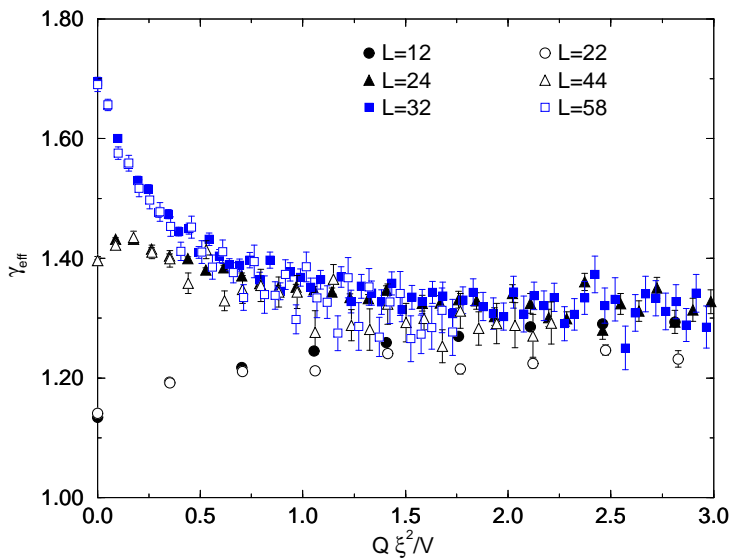}}
\vspace{-5mm}
\caption{Scaling test of $\gamma_{\it eff}$ for ${\rm CP}^{3}$FP with
$\beta=3.0$ (solid symbols) and $\beta=3.4$ (open symbols).}
\label{fig:CP3geffScal}
\end{figure}
\begin{figure}
\vspace{-5mm}
     \epsfxsize=8.4cm
     \centerline{\epsfbox{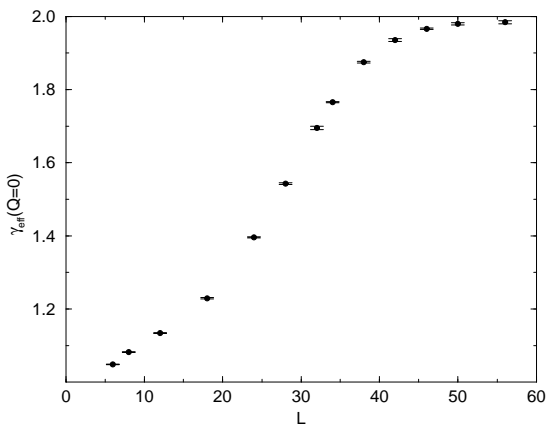}}
\vspace{-5mm}
\caption{Effective power $\gamma_{\it eff}$ at the origin as a function of the
lattice size $L$ for ${\rm CP}^{3}$FP at $\beta=3.0$.}
\label{fig:CP3geffQ0}
\end{figure}
\begin{figure}
\vspace{-10mm}
     \epsfxsize=12cm
     \centerline{\epsfbox{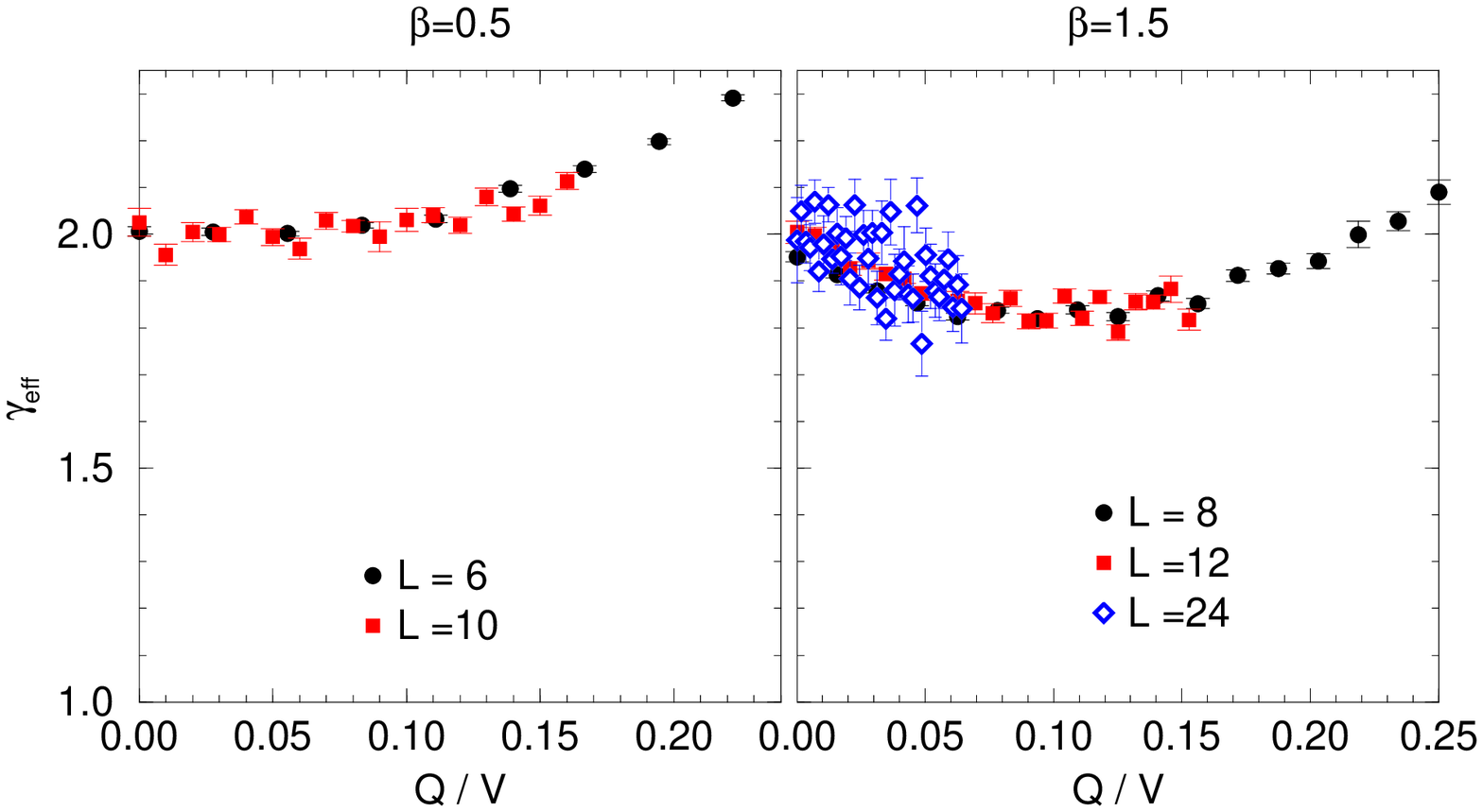}}
\vspace{-15mm}
\caption{$\gamma_{\it eff}$ for ${\rm CP}^{3}$FP at strong couplings.}
\label{fig:CP3geffStrong}
\end{figure}

In Fig.~\ref{fig:CP3geffScal} we compare the results for  $\gamma_{\rm
eff}$ at the two weak couplings, $\beta=3.0$ and 3.4. In order to make this
comparison,  we have to plot the data as a function of the physical 
charge density
instead of the filling fraction. By doing this, the data at the two couplings
collapse to a common line, showing scaling also for $\gamma_{\it eff}$. This
finding also confirms the approach from different sides to the asymptotic
value for different physical lattice sizes.
Let us finally focus on the value of $\gamma_{\it eff}$ obtained at the
origin, $Q=0$. As observed in Fig.~\ref{fig:CP3geff}, it takes values
between 1 and 2 and exhibits strong dependence on the lattice size. This
lattice size dependence is shown in Fig.~\ref{fig:CP3geffQ0}. We clearly
observe a crossover from values close to 1 for small lattices to values
approaching 2 in the limit of large lattices. In this way, we recover
a Gaussian charge distribution at weak coupling. It is realized, however,
only in the limit of large volume and only near vanishing topological
charge.
The effect observed at weak coupling is also observed for strong coupling, 
as shown in
Fig.~\ref{fig:CP3geffStrong}. Here we find $\gamma_{\it eff}$ to be
around 2, in accordance with the observation of Gaussian behavior at
strong coupling in Refs.~\cite{rf:HITY} and \cite{rf:IKY}. However,
if the filling fraction becomes large, we again see an increase of 
$\gamma_{\it eff}$, so that even
values larger than 2 can be obtained.

It is interesting to note in connection to this behavior that numerical 
simulations in
the strong coupling region, where a Gaussian distribution was observed
in previous studies,~\cite{rf:HITY,rf:IKY} are confined at the
same time in a region
of large volumes {\em and} in a region of small physical charge
density. The reason for this is that for a strong coupling, the correlation
length $\xi$ is small.  The lattice size $L$, however, should not be chosen
to be small, so that the lattice can contain more than just a few
instantons.  As a result, the ratio $L/\xi$ becomes large, which means that
we are in a region of large physical volumes. Moreover, even though in the
above-mentioned previous calculations the range of $Q$ was taken up to
fairly large values, they still correspond to a small physical
charge density $\xi^2 Q/V$, because  values of $L/\xi$ are large.
\section{Numerical results for ${\rm CP}^{1}$FP and ${\rm CP}^{3}$ST}
\label{sec:CP1results}
It is  worthwhile to investigate whether scaling behavior is 
observed, and if not, to what extent it breaks down for other cases of 
the ${\rm CP}^{N-1}$ model. 
For  comparison,  we considered a model with  
${\rm CP}^{1}$ FP action.
 As a parametrization of the ${\rm CP}^{1}$FP action, we used the  
parametrization listed in Table 4 of the original paper by Hasenfratz 
and Niedermayer,~\cite{rf:HN} in which 24  local coupling constants are 
employed. 
For updating configurations, we used the Metropolis algorithm. 
Typically, we performed one to several million sweeps per set, depending on the 
coupling constant $\beta$ and $L$. For the set method described in 
section~\ref{sec:Form}, we always chose 4 bins for each set and the trial 
function 
to be Gaussian. The values of $\beta$ ranged from 0.7 to 1.1 and the
lattice sizes from 12 to 62. \par
\begin{figure}
\vspace{-2mm}
    \epsfxsize=7.5cm
    \centerline{\epsfbox{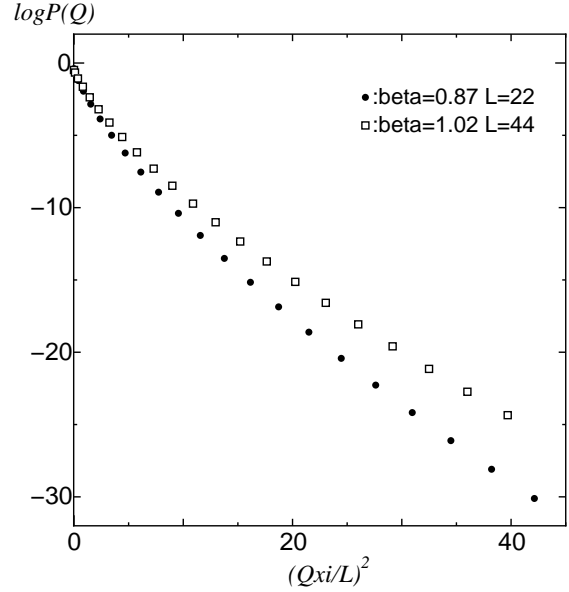}}
\vspace{-2mm}
\caption{Topological charge distribution $\log P(Q)$ vs $(Q\xi(\theta=0)/L)^{2}$
	 for ${\rm CP}^{1}$FP.
($\beta$, $L$) = (0.87,22) corresponds to the solid dots and (1.02,44) to the 
 open boxes.}
\label{fig:CP1PQ}
\end{figure}

\begin{figure}
\vspace{-3mm}
    \epsfxsize=7.5cm
    \centerline{\epsfbox{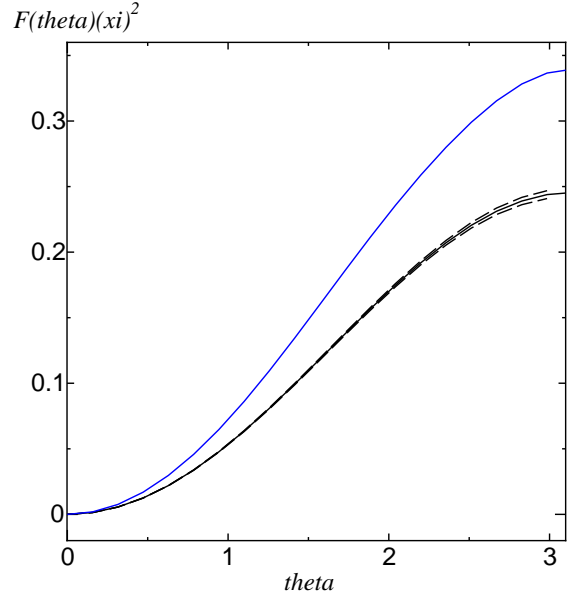}}
\vspace{-3mm}
\caption{Physical free energy density  $F(\theta)\xi^{2}$ for ${\rm CP}^{1}$FP.
The upper curve corresponds to  ($\beta$, $L$) = (0.87,22) and the lower 
to (1.02,44).}
\label{fig:CP1free}
\end{figure}

\begin{figure}
\vspace{-5mm}
    \epsfxsize=7.5cm
    \centerline{\epsfbox{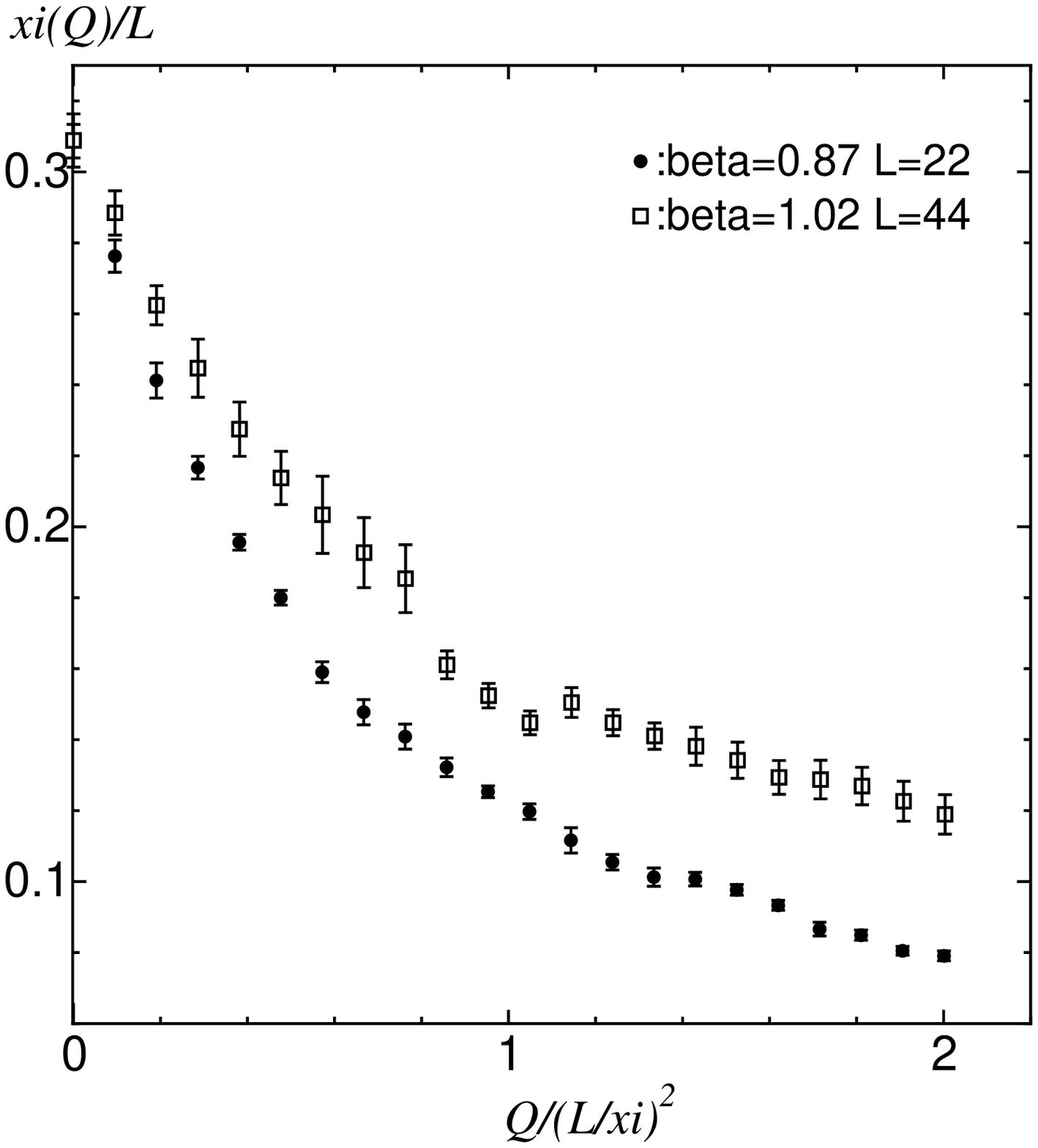}}
\vspace{-2mm}
\caption{The correlation length $\xi/L$ as a function 
  of  $Q/(L/\xi(\theta=0))^{2}$.
($\beta$, $L$) = (0.87,22) corresponds to the solid dots and (1.02,44) to the 
 open boxes.}
\label{fig:CP1xi}
\end{figure}

  
In order to study the scaling behavior, we 
compare the results for pairs  of parameter doublets  $(\beta, L)$, 
as  done in the 
 previous section,  so that each one of the pair  is  chosen to 
     have approximately the same physical volume, $(L/\xi(\theta=0))^{2}$.  
  Figures \ref{fig:CP1PQ} and \ref{fig:CP1free} show $P(Q)$ and $F(\theta)$, respectively, for ($\beta$, $L$)=(0.87,22) 
  and (1.02, 44), where  $L/\xi(\theta=0)$  is measured  to be   3.2.   
  These data exhibit a large difference, in contrast to 
  the    ${\rm CP}^{3}$FP case depicted in Fig.~\ref{fig:CP3PQScal},
   and the scaling  is  found to be clearly broken.  
  In Fig.~\ref{fig:CP1xi} we compare the correlation lengths $\xi/L$ as a function 
  of the physical charge density  $Q/(L/\xi(\theta=0))^2$. 
  The two sets of  data for $\xi/L$  also exhibit strong scaling
  violation, and large 
  deviation($\approx 30\%$) is found for $Q/(L/\xi(\theta=0))^{2} \approx 0.7$. 
  It is noted that this behavior is also seen for other choices of pairs 
  with different values of $L/\xi(\theta=0)$, e.g., $(\beta,L)$
  =(0.7, 12) and (0.9,30), with  $L/\xi(\theta=0)$ being 3.6.\par 
  \par
\vskip 0.5cm
We have found for the ${\rm CP}^{1}$ model  that scaling  is strongly 
violated, although the FP action is used.  This is expected, since 
 it is well known that dislocations do harm seriously in the  ${\rm CP}^{1}$ model. 
 In our calculations, we have used only the FP action, and the additional 
 use of a  FP charge might be expected to produce better  results.
However,  we suspect that even if we had used the FP charge, the scaling nature  would not have been
 improved. This comes from the observation 
  that even after the dislocations are eliminated by  a combined use of  the FP charge and 
   the FP action, 
      the scaling behavior of the lattice topological susceptibility 
    $\chi_{t}$ is strongly violated.~\cite{rf:DFP,rf:BBHN} 
     \par
We have also calculated the ${\rm CP}^{3}$ standard action for different choices of $\beta$ and $L$. We varied lattice sizes from 18 to 54 and $\beta$ from 
1.8 to 2.3 and looked for pairs of doublets $(\beta,L)$ such that $L/\xi$ 
for each pair is approximately the same.  Values of $L/\xi$ were chosen
to be from about 2.7 up to 4.2, which are similar to the values used for
CP$^{1}$FP and CP$^{3}$FP. 
For updating configurations, 
we used the Metropolis algorithm, and the number of sweeps used for each
set was approximately same as for CP$^{1}$FP.
Use of the standard action for the  ${\rm CP}^{3}$ model exhibits strong
violation, similar to the case 
 ${\rm CP}^{1}$FP, for $P(Q)$ and $F(\theta)$, as well as the correlation
length $\xi/L$. This is easily understood, since the standard action is 
very poor in representing continuum physics. The behavior of $\xi/L$ 
exhibits somewhat small deviation compared to that for CP$^{1}$FP.

\section{Debye-H\"uckel approximation}
\label{sec:DH}

In Refs. \cite{rf:HITY} and \cite{rf:IKY} it has been shown that a first-order
 phase
transition exists at $\theta=\pi$ when the topological 
charge distribution $P(Q)$ is Gaussian, and its volume dependence is
 given by $P(Q)\sim \exp[-\alpha Q^{2}/V]$, where $\alpha$ is a constant 
that depends on 
the coupling constant $\beta$. Such a Gaussian charge distribution has been
found in the region of very strong coupling. As $\beta$ becomes larger for
some fixed volume, $P(Q)$ has been observed to deviate  
from the Gaussian form, and thus the first-order phase transition gradually 
disappears. We note that $P(Q)$ for a system that consists of instantons 
and anti-instantons obeying the Poisson distribution also behaves like 
a Gaussian for values of the parameter corresponding to the strong
 coupling region. Therefore, we expect instantons and 
anti-instantons in such a system to behave like a dilute gas in 
the strong coupling 
region. It is thus of interest to determine the nature of the dynamics 
of instantons displayed by a system for which 
$P(Q)$ is not Gaussian. In order to investigate this issue, we use the 
Debye-H\"uckel model,~\cite{rf:DM} which is based on an instanton 
quark picture,~\cite{rf:FFS} and in which correlations 
between particles (instanton quarks) are weak.

\subsection{Instanton quark picture and Debye-H\"uckel model}

In this subsection we explain the concept of the instanton quark picture 
and the
Debye-H\"uckel model (D-H model). In order for this paper to be
self-contained, we also give a summarized overview of the results obtained in
previous works.~\cite{rf:DM,rf:FFS} 
In Ref.~\cite{rf:FFS}, Fateev, Frolov and Schwarz analyzed 
Euclidean Green's functions and the partition function to investigate how instantons
affect the dynamics of the CP$^{N-1}$ model. The partition function of this
model in the continuum is defined by
\begin{equation}
	Z=\int{\cal D}z{\cal D}\overline{z}\prod_{x}\delta(
	   \sum_{\alpha=1}^{N}|z_{\alpha}(x)|^{2}-1) e^{-S},
\end{equation}
and the action is defined by
\begin{equation}
	S=\beta\int d^{2}x\sum_{\alpha=1}^{N} \{|\partial_{\mu}
		 z_{\alpha}(x)|^{2}+(\overline{z}_{\alpha}(x) 
		 \partial_{\mu}z_{\alpha}(x))^{2} \},
\end{equation}
where $\overline{z}_{\alpha}$ denotes the complex conjugate of $z_{\alpha}$.
The $(N-1)$-dimensional complex projective space is defined by introducing 
a field as
\begin{eqnarray}
	{\it w}_{\alpha}(\zeta)=z_{\alpha}(\zeta) \big/ z_{N}(\zeta) & ; \;
			    \alpha=1,2,\cdots,N-1 \; &, \mbox{where } 
				z_{N}(\zeta)\neq 0,
\end{eqnarray} 
in the complex plane $\zeta=x+iy$. As a next step, the action is rewritten 
in the form
\begin{equation}
	S=\beta\int d^{2}x \frac{1}{1+\rho^{2}}\Bigl[ \sum_{\alpha}
	 |\partial_{\mu}{\it w}_{\alpha}|^{2}-\frac{\sum_{\alpha}
	 \overline{{\it w}}_{\alpha}\partial_{\mu}{\it w}_{\alpha}
	 \sum_{\gamma}{\it w}_{\gamma}\partial_{\mu}\overline{{\it w}}
	 _{\gamma}}{1+\rho^{2}}\Bigr],
\end{equation}
where $\rho^{2}=\sum_{\alpha=1}^{N-1}|{\it w}_{\alpha}|^{2}$ and $d^{2}x=\frac{1}{2}d\zeta d\overline{\zeta}$.
A general $q$-instanton solution is given by 
\begin{equation}
	z_{\alpha}(\zeta)=c_{\alpha}\prod_{i=1}^{q}(\zeta-a_{\alpha}^{i}), 
\end{equation}
where $c_{\alpha}$ and $a_{\alpha}^{i}$ are complex parameters. 
The superscript $i \> (\,i=1,2,\cdots,q)$ labels instantons, and the
subscript $\alpha\> (\, \alpha=1,2,\cdots,N)$ indicates the degrees of 
freedom of the field $z_{\alpha}$. 
The authors of Ref.~\cite{rf:FFS} investigated how the system behaves when the field fluctuates around 
the $q$-instanton solution~\cite{rf:Jev,rf:For} and showed that the partition function 
takes the form
\begin{equation}
	  Z=\sum_{q}W^{q}(q!)^{-N}\int\prod_{\alpha=1}^{N}
	    \prod_{i=1}^{q}d^{2}c_{\alpha}d^{2}a_{\alpha}^{i} 
	    \exp\bigl\{ -{\cal H}_{q}(a,c)/T\bigr\} \delta
	    \bigl(\sum_{\alpha=1}^{N}|c_{\alpha}|^{2}-1\bigr)
		\label{eqn:part},
\end{equation}
where $W$ is a constant that depends on the topological charge, and
$T=1$. ${\cal H}_{q}$ in Eq.~(\ref{eqn:part}) is given by
\begin{equation}
	{\cal H}_{q}(a,c)=\frac{N}{4\pi}\int d^{2}x \Bigl[\log\rho \partial^{2}
	   \log\rho-\log \bigl(\prod_{\alpha} |c_{\alpha}|^{2q}
	   \prod_{i>j}|a_{\alpha}^{i}-a_{\beta}^{j}|^{2})\Bigr].
		\label{eqn:hamil}
\end{equation}
Equation~(\ref{eqn:part}) can be interpreted as the partition function of a 
system of two-dimensional classical particles with interaction energy 
given by Eq.~(\ref{eqn:hamil}) in the grand canonical ensemble. 
The classical particles 
are at positions $a_{\alpha}^{i}$ and interact with each other. Note that 
$T$ plays the role of a temperature.

 In the particular case $N=2$, i.e.  for the ${\cal O}(3)$ 
non-linear sigma model, the grand partition function is given by
\begin{eqnarray}
	Z^{\prime}&=&\sum_{q}W^{q}(q!)^{-2}\int
	  \prod_{i}d^{2}a_{1}^{i} d^{2}a_{2}^{i}\exp\bigl\{ 
	  -{\cal H}^{\prime}_{q}(a)/T \bigr\}, \label{eqn:mod-part} \\
	  {\cal H}^{\prime}_{q}(a)&=&-\sum_{i<j}^{q}\log|a_{1}^{i}-a_{1}^{j}|
	  ^{2}-\sum_{i<j}^{q}\log|a_{2}^{i}-a_{2}^{j}|^{2}+\sum_{i,j}
	  \log|a_{1}^{i}-a_{2}^{j}|^{2}.
		\label{eqn:mod-hamil}
\end{eqnarray}
We see that in the Hamiltonian (\ref{eqn:mod-hamil}), a particle 
possesses a ``charge''$(\alpha=1,2)$ and that particles of equal charge 
interact repulsively, while those of opposite charge interact attractively. 
Thus we can interpret this model as a two-dimensional classical Coulomb system 
that consists of $2q$ particles with positive and negative charges.
Furthermore, if the locations 
$\{ a_{\alpha}^{i}; i=1,2,\cdots,q, \alpha=1,2\}$ satisfy the conditions 
$|a_{1}^{i}-a_{2}^{i}| \ll |a_{1}^{i}-a_{1}^{j}|,|a_{2}^{i}-a_{2}^{j}|  
(i \neq j)$, we can regard $\frac{1}{2}(a_{1}^{i}+a_{2}^{i})$ as the 
position and $\frac{1}{2}|a_{1}^{i}-a_{2}^{i}|$ as the size of the $i$-th 
instanton. Due to the interaction (\ref{eqn:mod-hamil}), a pair of 
particles with opposite charges tends to make up an instanton with neutral 
charge. For this reason, the particles are called ``instanton quarks''.~\cite{rf:FFS} 
\par
In the case of a $\overline{q}$-anti-instanton configuration,
 the solution is given by
\begin{eqnarray*}
	z_{\alpha}(\zeta)=\overline{c}_{\alpha}\prod_{i=1}^{\overline{q}}
	   (\overline{\zeta}-\overline{b}_{\alpha}^{i}), \; \alpha=1,2, 
\end{eqnarray*}
and the partition function has the same form as in Eq.~(\ref{eqn:mod-part}) 
but with the variables $a_{\alpha}^{i}$ replaced by $b_{\alpha}^{i}$. 
In analogy with the ${\cal O}(3)$ non-linear sigma 
model, the CP$^{N-1}$ model can be interpreted as a classical system that 
consists of instanton quarks with ``multicolors'' $(\alpha=1,2,\cdots,N)$ 
such that instanton quarks with the same color interact repulsively and 
those with different colors interact attractively. 
\par
Recently Diakonov and Maul~\cite{rf:DM} investigated the effects of 
instantons and 
anti-instantons for the CP$^{N-1}$ model by using the instanton quark picture.
They proposed a configuration with $N_{+}$ instantons and $N_{-}$ 
anti-instantons by assuming a product ansatz of the form~\cite{rf:BuLi}
\begin{equation}
	z_{\alpha}(\zeta)=\prod_{i=1}^{N_{+}}
	  (\zeta-a_{\alpha}^{i})\prod_{j=1}^{N_{-}}
	   (\overline{\zeta}-\overline{b}_{\alpha}^{j}),
	\label{eqn:ansatz} \nonumber 
\end{equation}
where $\alpha=1,2,\cdots,N$. Since Eq.~(\ref{eqn:hamil}) cannot be 
analytically solved for $N>2$, they simplified the multi-body interaction 
into an interaction of the two-body type. 
Furthermore, they applied the
Debye-H\"uckel approximation to  
this model; i.e., they assumed that the correlations between instanton 
quarks are very weak and individual quarks interact with a mean field. 
The partition function turns out to be
\begin{equation}
	Z(\theta)=\sum_{N_{+},N_{-}}Z(N_{+},N_{-})\exp\{i\theta(N_{+}-N_{-})\},
	\label{eqn:Ztheta}
\end{equation}
where the partition function $Z(N_{+},N_{-})$ is given by
\begin{eqnarray}
	Z(N_{+},N_{-})&=&\exp\Biggl[-\frac{N^{2}V_{d}}{4(N-1)}
	 \beta\biggl\{\frac{
	 \tilde{n}_{+}+\tilde{n}_{-}}{2}\Bigl(\log\frac
	 {\tilde{\Lambda}^{2}}{4\pi\beta\sqrt{\tilde{n}_{+}
	 \tilde{n}_{-}(1-\beta^{\prime2})}}+1\Bigr) \nonumber \\
		 &&-\frac{1}{4}\sqrt{(\tilde{n}_{+}
	 -\tilde{n}_{-})^{2}+4\tilde{n}_{+}\tilde{n}_{-}
	 \beta^{\prime2}}\log\frac{\kappa_{+}}{\kappa_{-}}\biggr\}\biggr]
	 \frac{(V_{d}\Lambda^{2})^{N(N_{+}+N_{-})}}
	 {(N_{+}!)^{N}(N_{-}!)^{N}}. 		\label{eqn:zet}\nonumber  \\
\end{eqnarray}
Here
\begin{eqnarray*}
	\beta&=&1/T \\
	\tilde{n}_{\pm}&=&\frac{N}{N-1}\frac{N_{\pm}}{V_{d}} \\
	\kappa_{\pm}&=&2\pi\beta\big[(\tilde{n}_{+}+\tilde{n}_{-})
	 \pm\sqrt{(\tilde{n}_{+}-\tilde{n}_{-})^{2}+4\tilde{n}_{+}
	 \tilde{n}_{-}\beta^{\prime2}}\bigr] \\
	\tilde{\Lambda}&=&2\Lambda e^{\gamma_{E}}.
\end{eqnarray*}
In the above, $\beta^{\prime}$ is a coupling constant between 
instanton quarks  and anti-instanton quarks  
introduced by the ansatz represented by Eq.~(\ref{eqn:ansatz}), and $V_{d}(\equiv L_{d}^{2})$ 
is the volume of the system in the continuum. $\Lambda$ is 
a parameter like $\Lambda_{QCD}$, and  $\gamma_{E}$ is the Euler number. 
Note that $T$ plays the role of a temperature ( here $T=1$).
\par
We calculate $P(Q)$ from Eq.~(\ref{eqn:zet}) as
\begin{eqnarray}
	Z(\theta)&& =\sum_{N_{+},N_{-}}Z(N_{+},N_{-})
		\exp\{i\theta(N_{+}-N_{-})\}	\nonumber \\
	       && =\sum_{Q}\sum_{N_{+},N_{-}}\delta_{N_{+}-N_{-},Q}
		  Z(N_{+},N_{-})
	          \exp\{i\theta(N_{+}-N_{-})\} \nonumber \\ 
	       	&& \equiv \sum_{Q}\exp\{i\theta Q\}P(Q),
\end{eqnarray}
where
\begin{equation}
	P(Q)\equiv \sum_{N_{+},N_{-}}\delta_{N_{+}-N_{-},Q}Z(N_{+},N_{-}).
		\label{eqn:pqD} \nonumber
\end{equation}
 We carefully consider the behavior of $P(Q)$ in the next subsection.

\subsection{Analysis of P(Q) in terms of the instanton picture based on
instanton quarks}
\begin{figure}
\begin{center}
\epsfxsize=8.5cm
\epsfbox{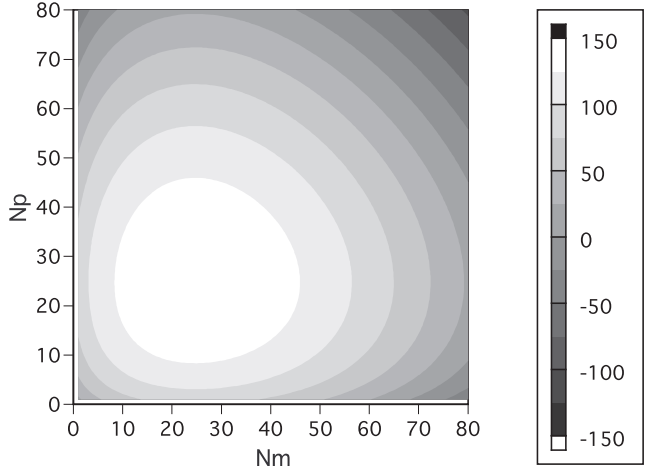}
\caption{Two-dimensional plot of the partition function $\log Z(N_{+},N_{-})$
	  for N=4, $L_{d}=4$, $\beta=1.0$ and $\beta^{\prime}=0.0$. Only 
	  results with $1 \leq N_{p},N_{m}\leq 80$ are plotted. Here 
          $N_{p}$ and $N_{m}$ are the number of instantons and 
          anti-instantons, respectively.}
\label{fig:DH-Z-2D}
\end{center}
\end{figure}

We now discuss the topological charge distribution $P(Q)$ in 
terms of instantons (anti-instantons). Before referring to $P(Q)$, 
let us first consider the behavior of $Z(N_{+},N_{-})$. 
In Fig.~\ref{fig:DH-Z-2D} we give a
two-dimensional plot of $Z(N_{+},N_{-})$
obtained from Eq.~(\ref{eqn:zet}) for $L_{d}=4$, $\beta=1.0$ and 
$\beta^{\prime}=0.0$. These values of the parameters were chosen 
to be the same as those in Ref.~\cite{rf:DM}. From this point,  we set 
$\Lambda$ to unity.~\cite{rf:DM} 
Figure~\ref{fig:DH-Z-2D} displays typical behavior of $Z(N_{+},N_{-})$. 
This function is symmetric under interchange of $N_{+}$ and 
$N_{-}$ and has a maximum at some non-zero value of $N_{+}(=N_{-})$. 
The region around the maximum gives a dominant contribution to the charge 
distribution $P(Q)$ at $Q=0$. It is interesting to see that 
a fairly large number of instantons $(N_{+}=N_{-}\approx 30)$ contribute to 
$P(0)$. 
\begin{figure}
\begin{center}
\input epsf
\epsfxsize=7cm
\epsfbox{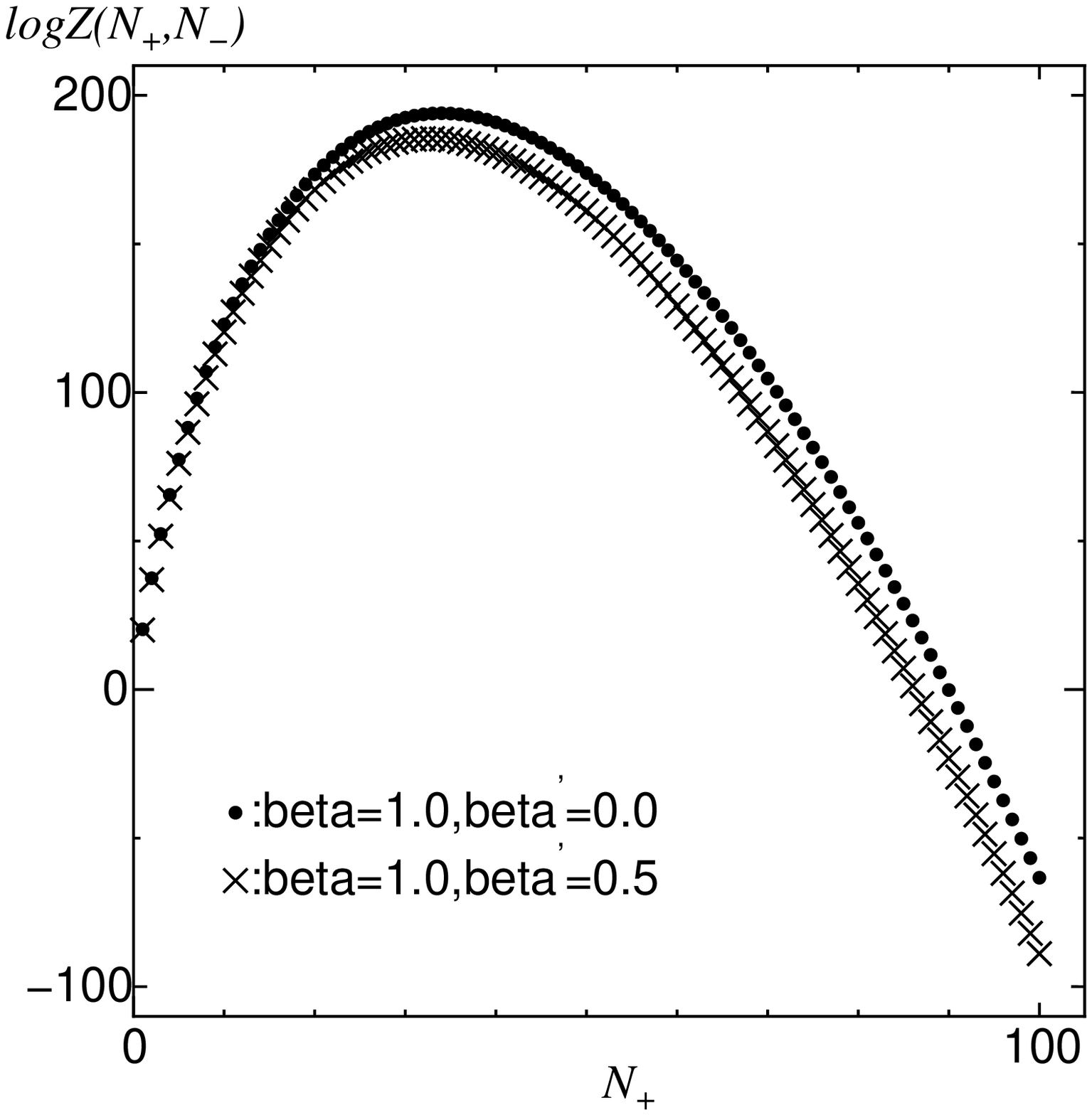}
\caption{The logarithm of $Z(N_{+},N_{-})$ for the case Q=0 with
$\beta=1.0$ and $L_{d}=4$. The results for $\beta^{\prime}=0.0$ are 
represented by solid dots and the results for $\beta^{\prime}=0.5$ by crosses.}
\label{fig:DH-Z-betaP}
\vskip 0.5cm
\epsfxsize=7cm
\epsfbox{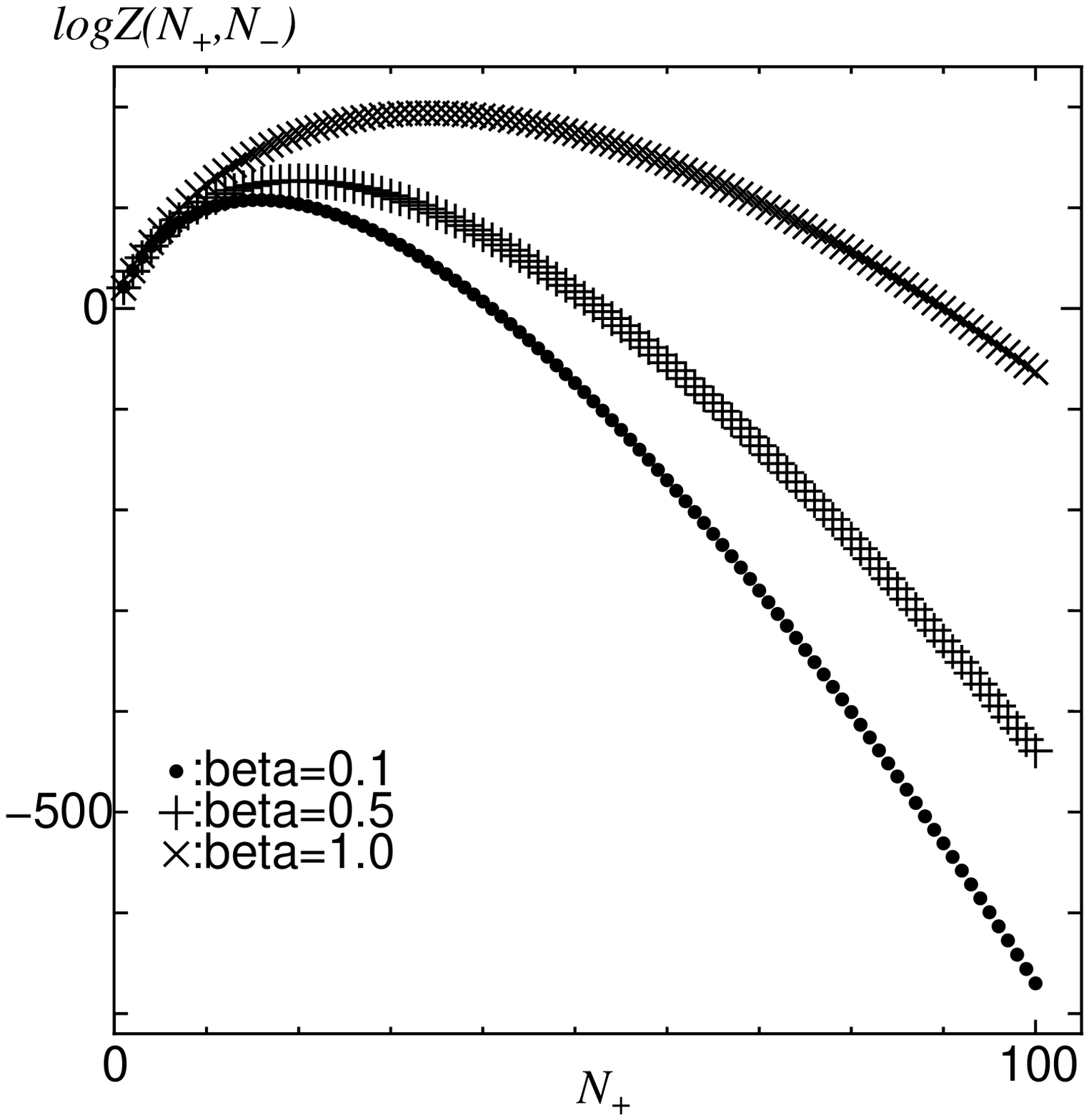}
\caption{The logarithm of $Z(N_{+},N_{-})$ in the case Q=0 with $L_{d}=4$ 
	  and $\beta^{\prime}=0.0$. $\beta$ is chosen as 0.1($\bullet$), 
	  0.5($+$) and 1.0($\times$).}
\label{fig:DH-Z-beta}
\end{center}
\end{figure}
In Fig.~\ref{fig:DH-Z-betaP} we compare $Z(N_{+},N_{-})$ 
for two different values of $\beta^{\prime}$ ($\beta^{\prime}=0.5$ and 
$\beta^{\prime}=0.0$) in the case $Q=0$ $(N_{+}=N_{-})$. 
We find that $Z(N_{+},N_{-})$ has only weak $\beta^{\prime}$ dependence. 
In fact, this is true for other choices of the parameters as well. 
This result shows that the correlations between instanton quarks  and 
anti-instanton quarks 
can be ignored.~\cite{rf:DM} For this reason, we  set 
$\beta^{\prime}=0$. 
Figure~\ref{fig:DH-Z-beta} shows how $Z(N_{+},N_{-})$ depends on $\beta$ 
for a fixed volume. We see that as  $\beta$ becomes larger, 
the position of the maximum of $Z(N_{+},N_{-})$ shifts to the right and 
configurations with larger numbers of  
instantons and anti-instantons are generated. Note that $\beta=1.0$ corresponds
 to a physical system.
For fixed $\beta$, on the other hand, the peak of $Z(N_{+},N_{-})$ shifts 
upward and to the right with increasing volume. This is intuitively
understandable. 

We are now in a position to study $P(Q)$. 
\begin{figure}
\begin{center}
\input epsf
\epsfxsize=7cm
\epsfbox{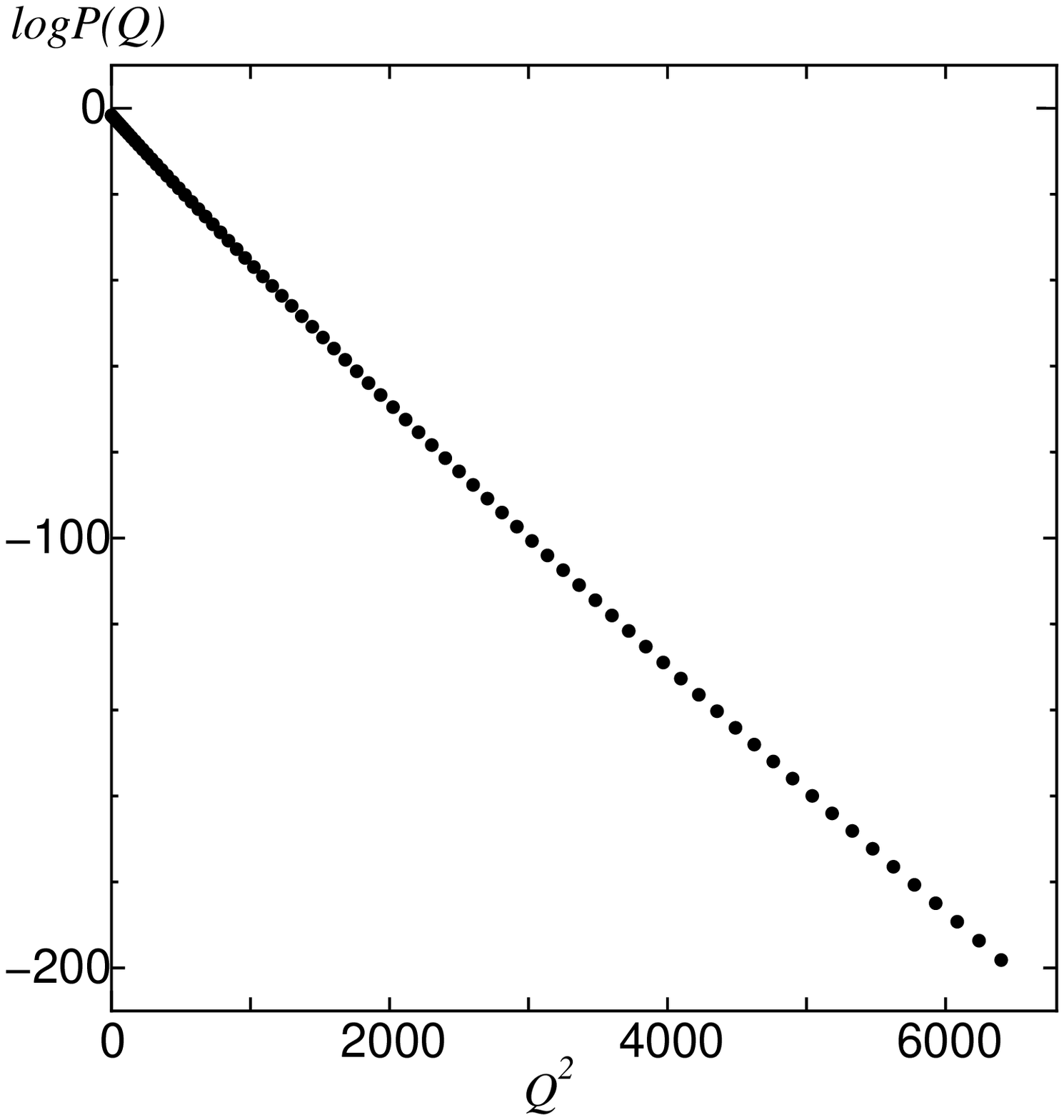}
\caption{The logarithm of $P(Q)$ as a function of $Q^{2}$ in the 
	  Debye-H\"uckel approximation for $\beta=1.0$, 
	  $\beta^{\prime}=0.0$ and $L_{d}=4$.}
\label{fig:DH-PQ}
\vskip 0.5cm
\epsfxsize=7cm
\epsfbox{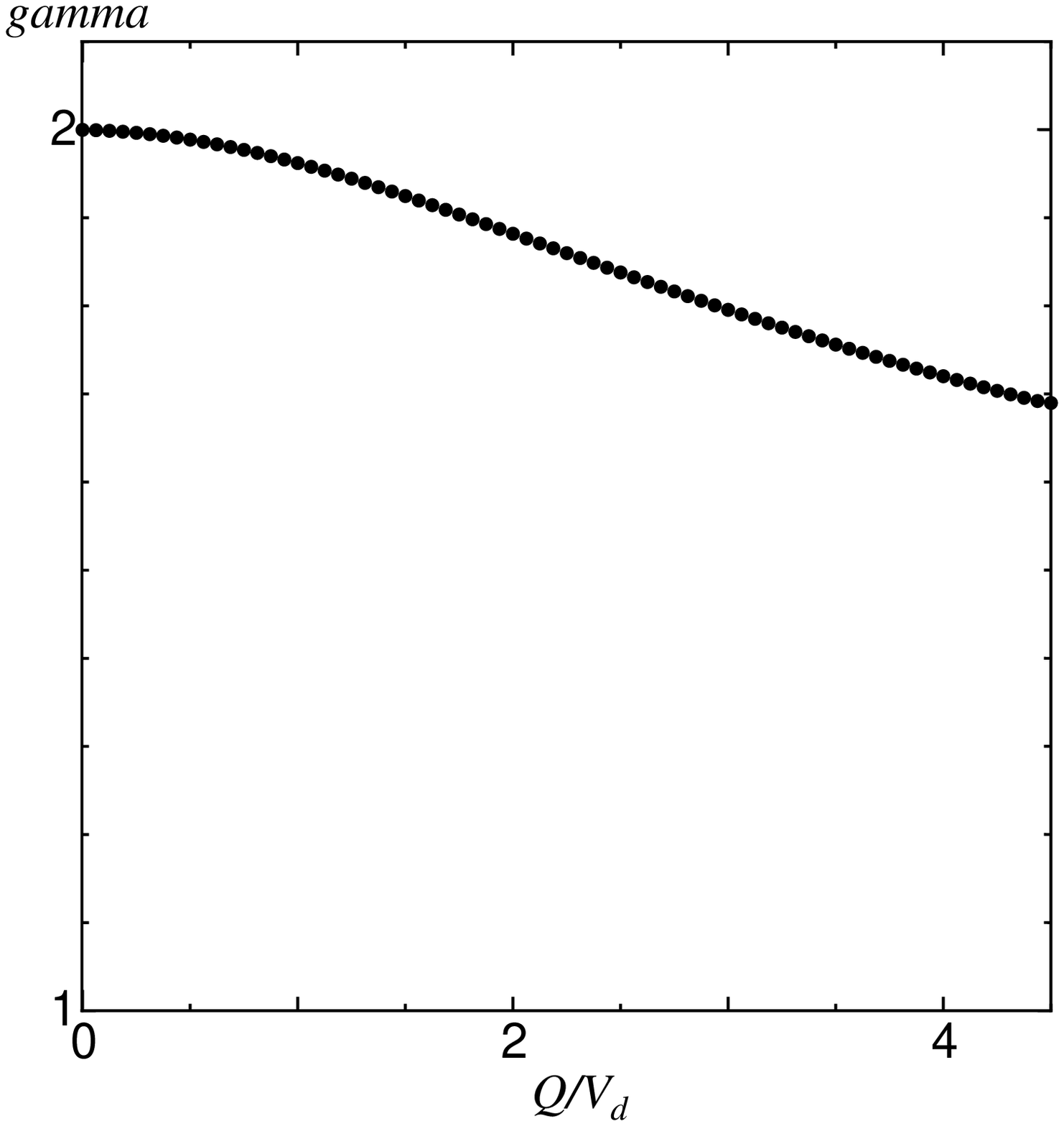}
\caption{$\gamma_{\it eff}^{(D)}$ obtained from $P(Q)$ for $\beta=1.0$, 
	  $\beta^{\prime}=0.0$ and $L_{d}=4$.}
\label{fig:DH-gam-org}
\end{center}
\end{figure}
In Fig.~\ref{fig:DH-PQ} we plot $P(Q)$ obtained from $Z(N_{+},N_{-})$ 
for $\beta=1.0$ and $L_{d}=4$. At first sight, $P(Q)$ 
looks like a Gaussian.  
 In order to investigate the behavior of $P(Q)$ in more detail, we calculate 
the effective power $\gamma_{\it eff}$ introduced in 
subsection~\ref{sec:CP3GEFF}. 
We recall that in Monte-Carlo simulations, $\gamma_{\it eff}$ is 
nearly equal to 2.0 at $Q/V=0$ and decreases slowly with increasing 
charge for large volumes, while for small volumes, $\gamma_{\it eff}$ is near 
unity at $Q/V=0$ and increases rapidly with increasing charge. 
It is interesting to investigate the cause of this difference and
whether it can be accounted for by the D-H model.
In Fig.~\ref{fig:DH-gam-org} we display $\gamma_{\it eff}^{(D)}$ 
as a function of 
the physical charge density $Q/V_{d}$, where $\gamma_{\it eff}^{(D)}$ is 
obtained in the D-H approximation. Obviously $\gamma_{\it eff}^{(D)}$ is 
different from 2.0 (Gaussian). Even if we take $\beta^{\prime}\neq 0$, the 
results do not change significantly. Compared to Fig.~\ref{fig:CP3geff}, the
 D-H approximation is found to reproduce the behavior of $\gamma_{\it eff}$ 
for large volumes. \par
 Let us now turn to small volumes, for which we believe that finite size 
effects are relevant for $P(Q)$ on a lattice. 
Since instantons with a size smaller than the lattice spacing fall through 
the lattice, the number of instantons is bounded from above
($\equiv N_{max}^{(lat)}$) in finite volumes. Contrastingly, the D-H model 
is defined in the continuum, and for it, an infinite number of instantons 
can be placed in finite volume. 
In order to reproduce the lattice results with this model, it is
necessary to restrict the upper limit ($\equiv N_{max}$) to a finite
value in the 
summation $\sum_{N_{+},N_{-}}Z(N_{+},N_{-})$ in Eq.~(\ref{eqn:pqD}).
Let us assume that $N_{max}$ mimics $N_{max}^{(lat)}$.
\begin{figure}
\begin{center}
\input epsf
\epsfxsize=7.2cm
\epsfbox{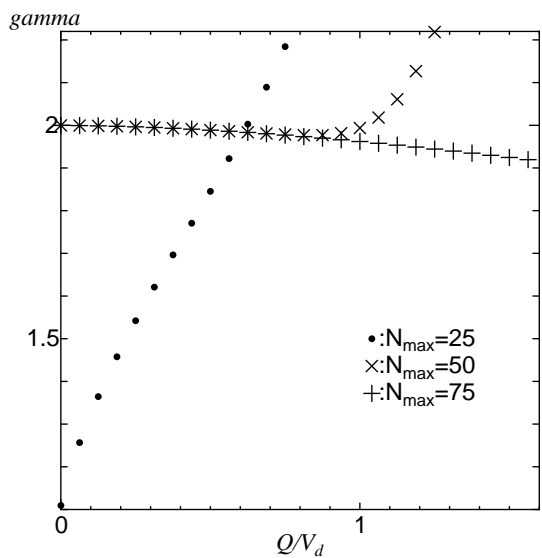}
\caption{Same as in Fig.~\ref{fig:DH-gam-org}, but with an upper limit 
	 $N_{max}$ in the summation 
	 $\sum_{N_{+},N_{-}}Z(N_{+},N_{-})$. $N_{max}$ is taken to be 
	  25($\bullet$), 50($\times$) and 75($+$).} 
\label{fig:DH-gam-cut}
\end{center}
\end{figure}
In Fig.~\ref{fig:DH-gam-cut} we display the results for $\beta=1.0$ 
and $L_{d}=4$ 
with three different choices of $N_{max}$. For large $N_{max}$ ($>40$), 
starting at $Q=0$, $\gamma_{\it eff}^{(D)}$ gradually decreases from $2.0$ 
and starts to diverge at some non-zero value of $Q/V_{d}$.
For smaller values of $N_{max}(=25)$, $\gamma_{\it eff}^{(D)}$ is close to 
unity at $Q=0$ and increases rapidly as a function of $Q$. At some small
value of the filling fraction, $\gamma_{\it eff}^{(D)}$ overshoots 2.0.
 The  behavior seen in Fig.~\ref{fig:DH-gam-cut} is quite similar to that
 seen in Fig.~\ref{fig:CP3geff}. It is noted that 
$\gamma_{\it eff}^{(D)}$ at $Q/V_{d}=0$ 
gradually shifts up from 1.0 up to 2.0 if we increase $N_{max}$ for small
 values, in accordance with the behavior seen in Fig.~\ref{fig:CP3geffQ0}. \par
We find  by varying  $N_{max}$ and by taking various values of 
$\beta$  and $L$ that when $N_{max}$ is chosen to be approximately equal
to or less than  the location of the peak of the distribution $Z(N_{+}, 
N_{-})$ (see Fig.~\ref{fig:DH-Z-2D}), $\gamma_{\it eff}^{(D)}$ 
 at $Q=0$ starts at a  value nearly equal 
to 1,  and when $N_{max}$ is larger than the location of the peak, 
 $\gamma_{\it eff}^{(D)}$ starts at 2. For the latter case,  
 $\gamma_{\it eff}^{(D)}$  slowly decreases as a function of $Q/V_{d}$, 
 and diverges at some value of  $Q/V_{d}$.  As $N_{max}$ becomes 
 even  larger, $\gamma_{\it eff}^{(D)}$ begins to diverge at a larger density 
 $Q/V_{d}$. 
These observations allow us to conclude that the behavior of $P(Q)$ 
found in Monte Carlo simulations reflects the dynamics of weakly correlated 
instanton quarks.

\subsection{Analysis of $P(Q)$ in terms of the Poisson distribution}

Before ending this section, we study $\gamma_{\it eff}$ obtained from 
a system that
consists of instantons and anti-instantons obeying the Poisson distribution. 
This is equivalent to a dilute gas system (DGA) of instantons and 
anti-instantons for which $P(Q)$ is given by 
\begin{equation}
	P(Q)=\sum_{N_{+},N_{-}}\frac{1}{2^{N_{+}+N_{-}}}\frac{(N_{+}+N_{-})!}
	   {N_{+}!N_{-}!}P_{N_{+}+N_{-}}\delta_{N_{+}-N_{-},Q}, 
		\label{eqn:pqP} \nonumber
\end{equation}
\[
	P_{n}\equiv \frac{\lambda^{n}}{n!}e^{-\lambda},
\]
\vspace{2.6mm}
where $\lambda/2$ is the average number of instantons (anti-instantons).

Equation~(\ref{eqn:pqP}) leads in the limit $\lambda \to \infty$ to
\begin{equation}
	P(Q)\sim \exp\bigl\{-\frac{1}{2\lambda}Q^{2}\bigr\}. 
	 \label{eqn:pqP-lim}
\end{equation}
With the identification $\lambda=\frac{V}{2\alpha}$,
 Eq.~(\ref{eqn:pqP-lim}) can 
be regarded as the distribution in the strong coupling region of MC 
simulations, where $\alpha$ is a constant that depends on $\beta$ appearing in 
$P(Q) \sim \exp[-\frac{\alpha}{V}Q^{2}]$ in such a manner  that $\alpha \to 0$ 
as $\beta \to 0$.

Let us study how $\gamma_{\it eff}$ behaves for $\lambda < \infty$.
\begin{figure}
\begin{center}
\input epsf
\epsfxsize=7.2cm
\epsfbox{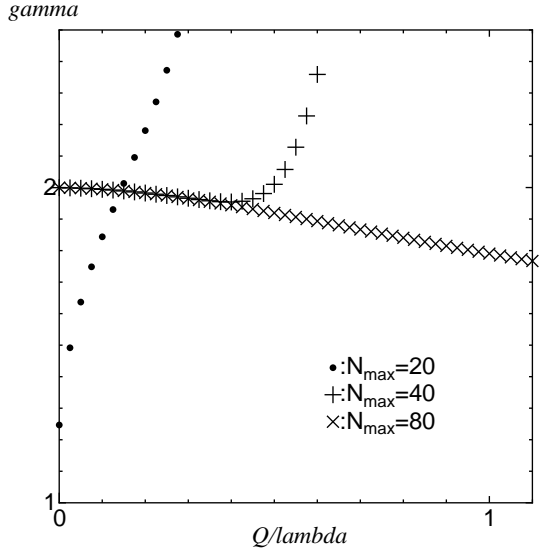}
\caption{$\gamma_{\it eff}^{(P)}$ for $\lambda=40$, where $\lambda$ is 
	 considered to be proportional to $V_{d}$.}
\label{fig:P-gam}
\end{center}
\end{figure}
Figure~\ref{fig:P-gam} displays 
$\gamma_{\it eff}(\equiv \gamma_{\it eff}^{(P)})$ 
as a function of $Q/\lambda$ for $\lambda=40$. 
 We find that the behavior of $\gamma_{\it eff}^{(P)}$ is very similar
 to that of $\gamma_{\it eff}^{(D)}$ for various values of $N_{max}$, 
where $N_{max}$ is the upper limit of the summation in Eq.~(\ref{eqn:pqP}).
Recalling that in 
the D-H model, correlations between {\it instanton quarks} are weak, 
while in the DGA there are no correlations between {\it instantons}, 
we can conclude that the D-H model also has little correlation between 
{\it instantons}.

\section{Conclusions and discussion}
\label{sec:Remarks}
\begin{itemize}
\item[(1)] 
   We have studied ${\rm CP}^{N-1}$ models with a topological term. In 
   order to obtain a description of continuum physics, we have employed 
   an FP action and 
   investigated scaling properties of  quantities such as $P(Q)$,  
   $F(\theta)$ and  the correlation length $\xi(Q)/L$. For ${\rm 
   CP}^{3}$FP we have observed good scaling behavior, while for  
   ${\rm CP}^{1}$FP and ${\rm CP}^{3}$ST we have found strong violations of 
   scaling.   
\item[(2)] 
   We have investigated the effective power $\gamma_{\it eff}$  of  $P(Q)$ 
    for CP$^{3}$FP. At a fixed value of $\beta$,  
   $\gamma_{\it eff}$ increases from 1.0 for small lattices as $Q$
   increases, while 
   $\gamma_{\it eff}$ decreases from 2.0 for large $L$. The values of 
   $\gamma_{\it eff}$ for both large and small lattices reach a common 
   line  at some small value of the filling fraction. 
   When finite size effects become significant, the value of 
   $\gamma_{\it eff}$ starts increasing beyond 2.0. 
\item[(3)] 
    We have studied the behavior of $\gamma_{\it eff}$ for the 
    analytical models, the Debye-H\"uckel approximation to an instanton 
    quark gas  
    of the CP$^{N-1}$ model, and the Poisson distribution of an  
    instanton gas. We have found that in these cases, 
    $\gamma_{\it eff}$ exhibits the same behavior as that   
    obtained in Monte Carlo simulations.  Finite size effects, which 
    emerge  by packing instantons into a finite volume, show up as an 
    increase beyond 2.0 of $\gamma_{\it eff}$. These observations 
    allow us to conclude that $P(Q)$ obtained in MC simulations 
    describes the dynamics of very weakly correlated instantons.  
\item[(4)] 
    Gaussian behavior of $P(Q)$ seems to be realized when volumes are 
    large {\it and } the physical charge density is small. In the strong 
   coupling region, these conditions are easily satisfied, because  
    correlation lengths are very short in this region. As a consequence,  
    there  exists a first-order phase transition at $\theta=\pi$. 
   In the weak 
    coupling region, on the other hand,  $\gamma_{\it eff}$ tends to 
   be 2.0 only at 
   vanishing topological charge when the volume increases. 
   The expectation value $\langle Q \rangle_{\theta}$ develops a peak, 
   and there is the tendency that the peak moves away from $\pi/2$ 
   towards $\pi$ as the volume increases. 
   However, we could not obtain  conclusive results about existence 
   of a  phase transition in  the infinite volume limit, because the 
   above conditions  are difficult to satisfy.      
\item[(5)] 
   When the volume is small, the value of $\gamma_{\it eff}$ increases 
   from 1.0 as a function of the charge in  all the cases 
   we have investigated,  not 
   only for simulations but also for analytical models.  It is an 
   interesting  question why $\gamma_{\it eff}$ is always bounded 
   from below by 1.0. 
   We think that perhaps this would be associated with some fundamental 
   property of probability theory. 
\item[(6)] 
   Although the analytical models can explain the behavior 
   of  $\gamma_{\it eff}$ qualitatively well,  $\gamma_{\it eff}$ for 
   small volumes increases more rapidly than in MC simulations. 
   This behavior might 
   be due to the sharp cut-off ($N_{max}$) used in the summation in 
   Eqs.~(\ref{eqn:pqD}) and ~(\ref{eqn:pqP}). We have tentatively smeared 
   the cut-off by 
   introducing a Gaussian function. As a result, we have observed 
   that the increase of $\gamma_{\it eff}$ becomes somewhat less rapid, 
   and it 
   tends to follow the common line before diverging. However,
   in order to draw a definite conclusion, a more systematic study is 
   needed.
\item[(7)]
  The fate  of the first-order phase transition at $\theta=\pi$ 
  in the strong coupling region is a relevant issue. As discussed in 
  subsection~\ref{sec:CP3FTH}, calculations of the 
  free energy and the expectation  value $\langle Q \rangle_{\theta}$ 
  have huge errors   already at $L=96$,  and thus we cannot address this
  issue. To increase lattice size while keeping errors reasonably small 
  requires an exponentially increasing number of sweeps. 
  We may need some novel algorithm to overcome this problem. 
\end{itemize}

\section*{Acknowledgments}
This work is supported in part by a Grant-in-Aid for Scientific Research (C) (2)
of the Ministry of Education (Nos. 11640248 and 11640250).  
Numerical simulations for CP$^{3}$FP were  performed on workstations 
at the
Center for Computational Physics,  University of Tsukuba,  and those for 
CP$^{1}$FP, CP$^{3}$ST and a part of CP$^{3}$FP were  done on workstations 
at  Saga University.




\begin{thebibliography}{99}
\bibitem{rf:tH} G. 't Hooft, \NP{B190 {\bf [FS3]},1981, 455}
     \\
\bibitem{rf:CR} J. L. Cardy and E. Rabinovici, \NP{B205 {\bf [FS5]},1982, 1}
     \\
	J. L. Cardy, \NP{B205 {\bf [FS5]},1982, 17}
     \\
\bibitem{rf:BDSL} G. Bhanot, R. Dashen, N. Seiberg and H. Levine, \PRL{53,1984, 
	519}
     \\
\bibitem{rf:Wie} U.-J. Wiese, \NP{B318,1989, 153}
     \\
	 W. Bietenholz, A. Pochinsky and U.-J. Wiese, \PRL{75,1995, 4524}
     \\ 
\bibitem{rf:HN} P. Hasenfratz and F. Niedermayer, \NP{B414,1994, 785}
     \\
\bibitem{rf:DFP}  M. D'Elia, F. Farchioni and A. Papa, \NP{B456,1995, 313}.
     \\
\bibitem{rf:BBHN} M. Blatter, R. Burkhalter, P. Hasenfratz and F. Niedermayer, 
	\PR{D53,1996, 923}
     \\
\bibitem{rf:Bu}  R. Burkhalter, \PR{D54,1996, 4121} 
     \\
\bibitem{rf:Sei} N. Seiberg, \PRL{53,1984, 637}
     \\
\bibitem{rf:HITY} A. S. Hassan, M. Imachi, N. Tsuzuki and H. Yoneyama, 
     \PTP{95,1995, 175}
     \\
	M. Imachi, S. Kanou and H. Yoneyama, \NP{B {\rm (Proc.Suppl)} 73,1999,
	 644}
     \\
\bibitem{rf:IKY} M. Imachi, S. Kanou and H. Yoneyama, \PTP{102,1999, 653} 
     \\
	R. Burkhalter, M. Imachi and H. Yoneyama, \NP{B {\rm (Proc.Suppl)} 83-84,2000, 562}
     \\ 
\bibitem{rf:DM} D. Diakonov and M. Maul, \NP{B571,2000, 91}
     \\
\bibitem{rf:FFS} V. A. Fateev, I. V. Frolov and A. S. Schwarz, \NP{B154,1979, 1}
     \\
	 V. A. Fateev, I. V. Frolov and A. S. Schwarz,
	   Sov.J.Nucl.Phys {\bf 30}(4),(1979), 590
     \\
\bibitem{rf:BeLu} B. Berg and M. L\"uscher, \NP{B190 {\bf [FS3]},1981, 412}
     \\
\bibitem{rf:BBCS} G. Bhanot, S. Black, P. Carter and R. Salvador, \PL{B183,1987, 
     331}
     \\
\bibitem{rf:KSC} M. Karliner, S. R. Sharpe and Y. F. Chang, \NP{B302,1988, 204}
     \\
\bibitem{rf:HM}  M. Hasenbusch and S. Meyer, \PR{D45, 1992, 4376}
     \\   
\bibitem{rf:Sch} S. Olejnik and G. Schierholz, \NP{B {\rm (Proc.Suppl)} 34,1994,
     709}
     \\
     G. Schierholz, \NP{B {\rm (Proc.Suppl)} 37A,1994, 203}
     \\
\bibitem{rf:PS} J. C. Plefka and S. Samuel, \PR{D56,1997, 44} 
     \\
\bibitem{rf:BK} W. Bock and J. Kuti, \PL{B367,1996, 242}
     \\
\bibitem{rf:Jev} A. Jevicki, \NP{B127,1977, 125}
     \\
\bibitem{rf:For} D. F\"orster, \NP{B130,1977, 38}
     \\
\bibitem{rf:BuLi} A. P. Bukhvostov and L. N. Lipatov, \NP{B180 {\bf [FS2]},1981, 116}
\end{thebibliography}
\end{document}